\documentclass[fleqn,usenatbib]{mnras}
\usepackage{newtxtext,newtxmath}

\usepackage{booktabs, makecell}
\usepackage{multirow}
\usepackage[T1]{fontenc}
\usepackage{ae,aecompl}
\usepackage[utf8]{inputenc}
\usepackage{graphicx}	% Including figure files
\usepackage{amsmath}	% Advanced maths commands
\usepackage{amssymb}% Extra maths symbols
\usepackage{amsbsy}
\usepackage{pdflscape}	% Landscape pages
\usepackage{lipsum}
\usepackage{adjustbox}
\usepackage{comment}
\title[Protocluster detection, using PCcones]{Protocluster detection in simulations of HSC-SSP and the 10-year LSST forecast, using PCcones}

\author[P. Araya-Araya et al.]{
Pablo Araya-Araya,$^{1}$\thanks{E-mail: paraya-araya@usp.br}
Marcelo C. Vicentin,$^{1}$
Laerte Sodr\'e Jr.,$^{1}$
\newauthor \hspace{0.01cm} Roderik A. Overzier$^{1,2}$ and Hector Cuevas$^{3}$
\\
$^{1}$Departamento de Astronomia, Instituto de Astronomia, Geof\'isica e Ci\^encias Atmosf\'ericas, Universidade de S\~ao Paulo \\ \hspace{0.01cm} Rua do Mat\~ao 1226 , Cidade Universit\'aria, 05508-900, S\~ao Paulo, SP , Brazil \\
$^{2}$Observat\'orio Nacional, Rua General Jos\'e Cristino, 77, S\~ao Crist\'ov\~ao, 20921-400, Rio de Janeiro, RJ, Brazil.\\ 
$^{3}$Departamento de F\'isica y Astronom\'ia, Universidad de La Serena, Av. Juan Cisternas 1200 Norte, La Serena, Chile.}
\date{Accepted XXX. Received YYY; in original form ZZZ}

\pubyear{2021}

\graphicspath{{Figures/}}
\begin{document}
\label{firstpage}
\pagerange{\pageref{firstpage}--\pageref{lastpage}}
\maketitle
\begin{abstract}
The progenitors of present-day galaxy clusters give important clues about the evolution of the large scale structure, cosmic mass assembly, and galaxy evolution. Simulations are a major tool for these studies since they are used to interpret observations. In this work, we introduce a set of “protocluster-lightcones”, dubbed PCcones. They are mock galaxy catalogs generated from the Millennium Simulation with the L-GALAXIES semi-analytic model. These lightcones were constructed by placing a desired structure at the redshift of interest in the centre of the cone. This approach allows to adopt a set of observational constraints, such as magnitude limits and uncertainties in magnitudes and photometric redshifts (photo-zs), to produce realistic simulations of photometric surveys. We show that photo-zs obtained with PCcones are more accurate than those obtained directly with the Millennium Simulation, mostly due to the difference in how apparent magnitudes are computed.
We apply PCcones in the determination of the expected accuracy of protocluster detection using photo-zs in the $z=1-3$ range in the wide-layer of HSC-SSP and the 10-year LSST forecast. With our technique, we expect to recover only $\sim38\%$ and $\sim 43\%$ of all massive galaxy cluster progenitors with more than 70\% of purity for HSC-SSP and LSST, respectively. 
Indeed, the combination of observational constraints and photo-z uncertainties affects the detection of structures critically for both emulations, indicating the need of spectroscopic redshifts to improve detection.
We also compare our mocks of the Deep CFHTLS at $z<1.5$ with observed cluster catalogs, as an extra validation of the lightcones and methods.
\end{abstract}
%%%%%%%%%%%%%%%%%%%%%%%%%%%%%%%%%%%%%%%%%%%%%%%%%%

%%%%%%%%%%%%%%%%% BODY OF PAPER %%%%%%%%%%%%%%%%%%
\section{Introduction}

The hierarchical model of structure formation \citep{blumenthal84, davis85, cole00} predicts protoclusters at high redshifts, i.e., systems which are the progenitors of the largest present-day virialized structures, the galaxy clusters. A formal definition of what a proper protocluster is has been suggested by \citet{rover}: any region at high redshift that will evolve into a galaxy cluster at $z=0$ or before. In practice, observations are only able to detect those protoclusters that are sufficiently overdense compared with their surroundings. 

Our knowledge of these structures can be guided by simulations, which can be used to predict the properties of the protoclusters themselves and their galaxy members. For instance, \citet{chiang1} have shown that a significant fraction of protocluster mass at $z=2.0$ is within a radius $\sim 4$ times larger than the typical virial radius of $z=0$ galaxy clusters. \citet{muldrew15} found that just $\sim 20 \%$ of the total protocluster mass is enclosed by the main dark matter halo at the same redshift. Investigating galaxy properties through the application of semi-analytic models, \citet{muldrew18} showed that galaxies in the principal halo reach their star formation peaks earlier than galaxies in smaller halos. These authors also showed that 80\% of the total stellar masses of protocluster members have already formed by $z = 1.4$, while for field galaxies, this amount is just 45\%. This occurs due to accelerated evolution in overdense regions. Recently, \citet{trebitsch20}, using the \texttt{OBELISK} hydrodynamical cosmological simulation, found that stellar populations provide enough energetic photons to complete the HI reionization at $z \sim 6$ without other additional ionizing sources (e.g., AGNs and/or collisions). These results may suggest that reionization happens from inside a galaxy overdense region out to the entire intergalactic medium.            

This decade has seen a significant progress in the study of galaxy clusters at high redshift. New protoclusters and high redshift cluster candidates  are being continuously reported from the analysis of surveys, mostly  in the optical and infrared. For example, \citet{chiang14} reported $36$ photometric redshift selected protoclusters at $z \sim 1.5 - 3.0$ in the COSMOS field. \citet{toshikawa16} cataloged $21$ protocluster candidates in the Deep survey of CFHTLS using the dropout technique; applying the same method over the wide layer of HSC-SSP area, \citet{toshikawa18} presented $210$ new candidates at $z \sim 4.0$. Also, \citet{martinache18} selected $2151$ and $228$  protocluster candidates from \textit{Planck} and \textit{Herschel} fields, respectively. From these, $89$ were observed with \textit{Spitzer}/IRAC, with more than $~92\%$ of them presenting significance overdensities. More recently, 
\citet{gonzalez19} have found a total of $1787$ high redshift cluster candidates in the sky covered by the Pan-STARRS and SuperCOSMOS surveys.

Another observational approach targets certain types of objects assuming that they probe dense regions. Examples include radio-loud active nuclei \citep[e.g.][]{over06, venemans07, hatch11a, hatch11b, hayashi12, wylezalek13, cooke14, hatch14, cooke16}, also found in simulations by \cite{orsi} and \cite{lovell}. At the same time, despite optically selected quasars being often used to trace overdense regions \citep[e.g.][]{boris, adams15, onoue18, stott20}, the population of quasars generally does not appear to probe proto-clusters \citep{champagne18, uchiyama18, vicentin20}. \cite{yoon19} suggest that massive galaxies are better tracers of overdense regions than quasars. Additionally, sub-millimeter observations reveal that many galaxies in protoclusters are gas-rich \citep{lacaille18, arrigoni18, noble19, cooke19}.   

A common approach to find high redshift structures is by adopting colour criteria using narrow-band or broad-band filters. With the former, it is possible to select H$\alpha$ or Ly$\alpha$ emitters (hereafter, HAEs and LAEs, respectively), and search for an excess of these objects at similar redshifts (e.g., \citet{shimakawa18a, shimakawa18b} for HAEs) and (e.g., \citet{venemans05, venemans07,chiang15, higuchi18} for LAEs). Similarly, \citet{toshikawa16, toshikawa18} found a large number of potential protoclusters from the clustering of Lyman Break Galaxies selected through the dropout technique. 

Spectroscopic follow-up is required to verify whether protocluster member candidates are indeed part of a protocluster. Membership confirmation is very challenging and expensive, since the sources are faint. 
However, the analysis of galaxy clustering in spectroscopic surveys designed to explore galaxy evolution can increase the number of real protoclusters. Examples include surveys such as z-COSMOS \citep{lilly07, lilly09}, the VIMOS Ultra Deep Survey (VUDS) \citep{lefevre15}, VANDELS \citep{mclure18}, among others. In particular, the future Prime Focus Spectrograph \citep{takada14} will be able to confirm a large number of protocluster candidates between $1.0 \lesssim z \lesssim 2.5$.

Some systems are expected to evolve into the biggest structures currently known in the local universe, making these protoclusters particularly interesting targets to study the first stages of galaxy assembly in dense regions \citep[e.g.][]{steidel98, pentericci00, cucciati14, miller18, cucciati18, shi19, long20}. 

Results from simulated data are commonly used to interpret observations \citep[e.g.][]{overzier09,chiang1, toshikawa16, lemaux18, jiang18, shi19}. In particular, lightcones are often constructed to analyze observational data from simulations that emulates observations \citep{blaizot05, kitzbichler07, merson13, overzier13, Stothert18}. The usage of lightcones helps to address the purity and completeness of detection or selection of structures \citep[Werner et al. in prep.]{kim16, ascaso16, costaduarte, euclid19, krefting20}, as long as observational constraints are properly taken in to account. 

Projects, such as the Vera Rubin Observatory's Legacy Survey of Space and Time (LSST), Euclid, and the James Webb Space Telescope, will open new windows to discoveries, and many new protoclusters and high-z clusters candidates should be found. Simulated data  has been used to evaluate  forecasts for these surveys \citep{bisigello, graham18, laigle19, graham20}. However, despite the usefulness of lightcones to explore different selection/detection criteria, some technical aspects of their construction may lead to difficulties in the future interpretations of observations. As an example, the number of massive (proto)clusters found in a specific redshift range is limited by the lightcone volume. 

Here we introduce the protocluster-lightcone, dubbed PCcones, as a tool to help in this type of analysis, focusing on photometric redshift surveys using broad-band photometry, and emphasizing structure detection at $1.0\leq z \leq 3.0$. It consists of a $\pi$ deg$^2$ lightcone with a pre-selected structure, like a $z=0$ galaxy cluster progenitor, placed at a desired higher redshift. This is particularly helpful for investigating protocluster candidate detections from imaging. Our motivation to introduce these PCcones is that we can add observational constraints to the analysis (like limits and errors in magnitudes) and examine their impact in the detection of galaxy overdensities, evaluating the likelihood of detected structures being real, determining detection rates, and estimating the expected quality of photometric redshift selection, allowing to estimate the probability that an observed overdensity for a given magnitude limit and photo-z selection is indeed a protocluster. This approach has several advantages, such as helping to design protocluster surveys, interpret their results, and, eventually, can also be useful to justify spectroscopic follow-ups. 

In this work, we assume a $\Lambda$CDM universe, with cosmological parameters equal to those obtained by the \textit{Planck1} mission results \citep{planck1}: $h = 0.673$, $\Omega_m = 0.315$ and $\Omega_{\Lambda} = 0.685$.

\section{Lightcone Construction}

In this section we describe the procedure adopted to construct protocluster lightcones with the Millennium Simulation \citep{springel05} and the \texttt{L-GALAXIES} semi-analytic model \citep{henriques15}. A main motivation for this choice is that the simulated  merger trees obtained with the \texttt{SUBFIND} algorithm \citep{springel01} are the basis of \texttt{L-GALAXIES} modeling. Additionally, the Millennium simulation has a dark matter particle mass ($m_p = 9.6 \times 10^{8}$ $M_{\odot}/h$) which allows modeling galaxies with $M_{\star} > 10^{8}$ $M_{\odot}/h$ with \texttt{L-GALAXIES}, and a box size ($L=480.279 $ Mpc$/h$) large enough to contain a reasonable number of massive (proto)clusters.

The results of this paper are obtained by placing, at different redshifts ($z=1.0$, $1.5$, $2.0$, $2.5$ and $3.0$), $20$ cluster progenitors  obtained from the Millennium Simulation with $ M_{z=0} > 1.37 \times 10^{14} M_{\odot}$ (see Section \ref{sec:pcones} for more details about the placed protoclusters); this value corresponds to low mass  Fornax-type protoclusters \citep{chiang1}. Here $M_{z=0}$ denotes the 
descendant cluster mass at $z=0$. We define a galaxy cluster following \citet{chiang1}, that is, a main Friend of Friends dark matter halo at $z=0$ with  $\texttt{M\_tophat} > 1\times 10^{14} \ M_{\odot}$\footnote{Notice that the $z=0$ snapshot in the \textit{Planck1} scaled catalog is not the $64th$ snapshot;  it is, instead, the $58th$.}.

\subsection{Synthetic Galaxies}\label{sec:maths}% used for referring to this section from elsewhere

The first step in the construction of lightcones is to obtain simulated galaxies. In this work, we used the \citet{henriques15} version of \texttt{L-GALAXIES} semi-analytic model.

In short, a semi-analytic model (SAM) corresponds to a set of differential equations describing the evolution of the baryonic components (the dark matter is simulated directly using a N-body code, in this case the Millennium Simulation); the model has free parameters that can be constrained by observations. The differential equations act over primordial gas associated to dark matter particles set at the beginning of a dark matter N-body simulation. The primordial gas evolves and is transformed into other baryonic components, such as stars, black holes, and intracluster hot gas, depending on the model. The SAM output  gives physical properties of the synthetic galaxies, such as stellar mass, gas mass, star formation rate, among others \citep[see ][for more details]{henriques15}. 

The most classical SAMs compatible with \texttt{SUBFIND} merger trees are \texttt{L-GALAXIES} \citep{croton06, delucia06, guo10, guo13, henriques15, henriques20} and \texttt{GALFORM} \citep{cole00, baugh05, bower06, gonzalez-perez14, lacey16, baugh19}. Although stellar mass functions and specific star formation rates are consistent with each other up to $z = 2.0$, there are differences in the implementation of the equations that describe galaxy formation and evolution \citep[see ][for a detailed description of their main differences]{guo16}. Some of them are the efficiency of AGN and stellar feedback, the definition of central and satellites galaxies, environmental processes, and the implementation of tidal stripping in \texttt{L-GALAXIES}, while this is not in \texttt{GALFORM}. This impacts the stellar mass functions of passive/star-forming and central/satellite galaxies and relations such as mass-metallicity and stellar mass-size. Our main motivation for using \texttt{L-GALAXIES} instead of \texttt{GALFORM} is the output of star formation history arrays. We use them to estimate the spectro-photometric properties of the model galaxies.

We apply the \texttt{L-GALAXIES} SAM to the Millennium Run simulation \citep{springel05} scaled to the cosmological parameters obtained from the \textit{Planck1} data \citep{planck1}; the scaling from the original cosmological parameters to the new ones has been performed using the \citet{angulo10} algorithm.

\subsection{Lightcone Space Definition}\label{sec:lightcone_space}

We want to place the progenitor of a galaxy cluster in the centre of a lightcone and at a redshift $z = z_p$.
We define the lightcone space as given by the orthonormal coordinates
$(\hat{u_1}, \hat{u_2}, \hat{u_3})$, similar to the procedure explained in \cite{kitzbichler07}. The line-of-sight direction $\hat{u_3}$ is set as $\vec{u_3} = (n,m,nm)$, and we adopt the values, $n=3$ and $m=4$.  Two additional orthonormal vectors, $\hat{u_1}$ and $\hat{u_2}$, are used to obtain the simulated angular coordinates: right ascension, $\alpha$, and declination, $\delta$ (see Section \ref{sec:spatial_coords}).

The progenitor of a galaxy cluster at $z=z_p$ resides in some snapshot of the Millennium Simulation. Since $z=z_p$ is not, in general, the redshift of a Millennium snapshot, we search for the protocluster in the snapshot $s_j$, with $z_j \leq z_p < z_{j+1}$, where $z_j$ and $z_{j+1}$ represent the redshift of the snapshots $s_j$ and $s_{j+1}$, respectively. 

\begin{figure}
    \centering
	\includegraphics[width=\columnwidth]{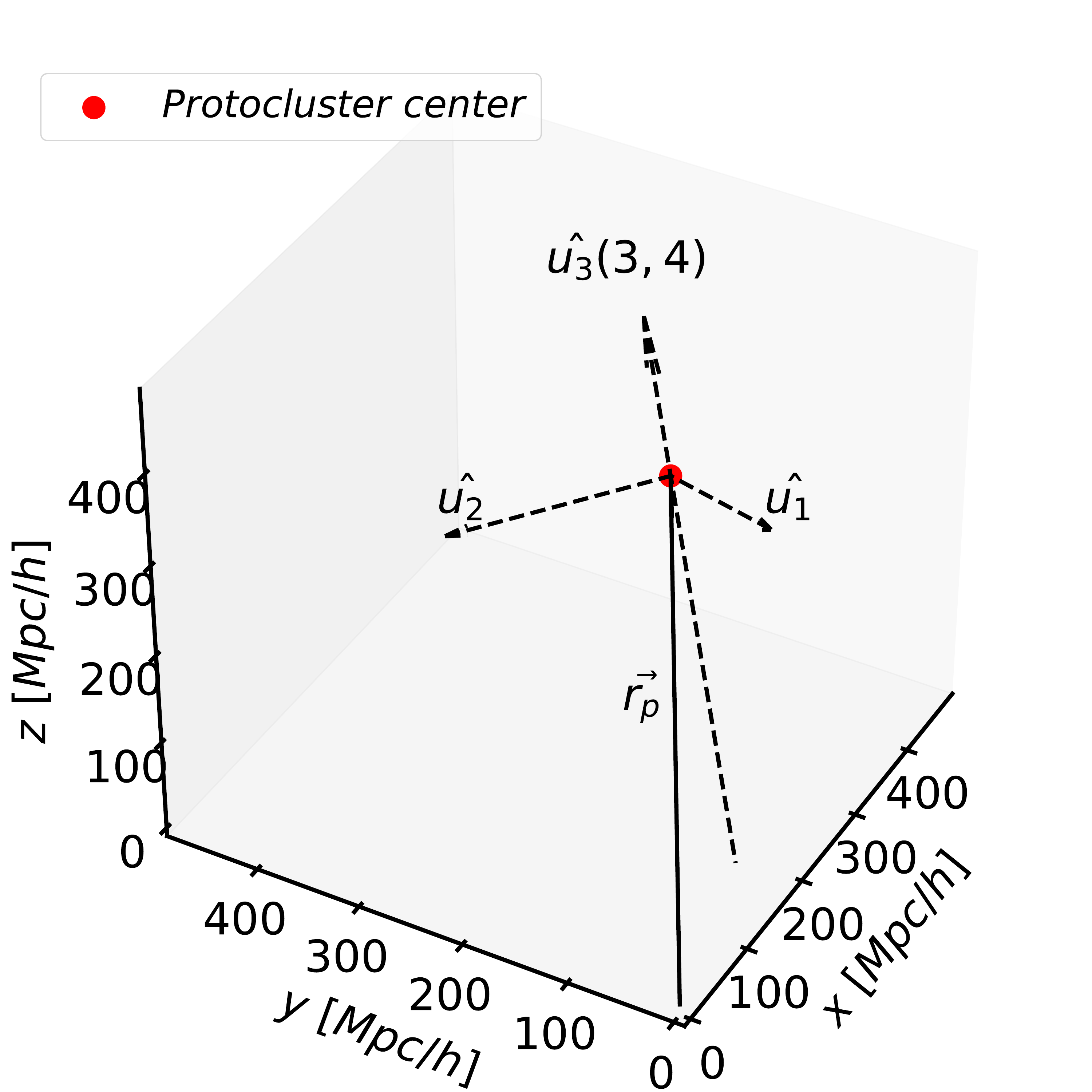}
    \caption{\small Definition of lightcone coordinates, where  $\vec{r_p}$ indicates the position of the protocluster that we want to place at $z_p$ in the lightcone. $\hat{u_3}$, $\hat{u_1}$, and $\hat{u_2}$ are orthogonal vectors that define the lightcone space. $\hat{u_3}(3,4)$ represents the line-of-sight with $n=3$ and $m=4$.}
    \label{direccion}
\end{figure}

To place the centre of mass of the protocluster, $\vec{r_p}$, at the comoving distance $d_C(z_p)$, we re-define the zero point, by putting it at the $z=0$ position. Then, if $\vec{r_p} = \hat{u_3} \ d_C(z_p) + \vec{r_o}$, we can establish the position of the zero point, $\vec{r_o}$, by:

\begin{equation}
    \vec{r_o} = \vec{r_p} - \hat{u_3}  d_C(z_p)
    \label{r_o}
\end{equation}
Figure \ref{direccion} shows an illustration of the lightcone space, with the line-of-sight, the orthonormal vectors, and the protocluster centre.

\subsection{Lightcone Volume}

\begin{figure}
    \centering
	\includegraphics[width=\columnwidth]{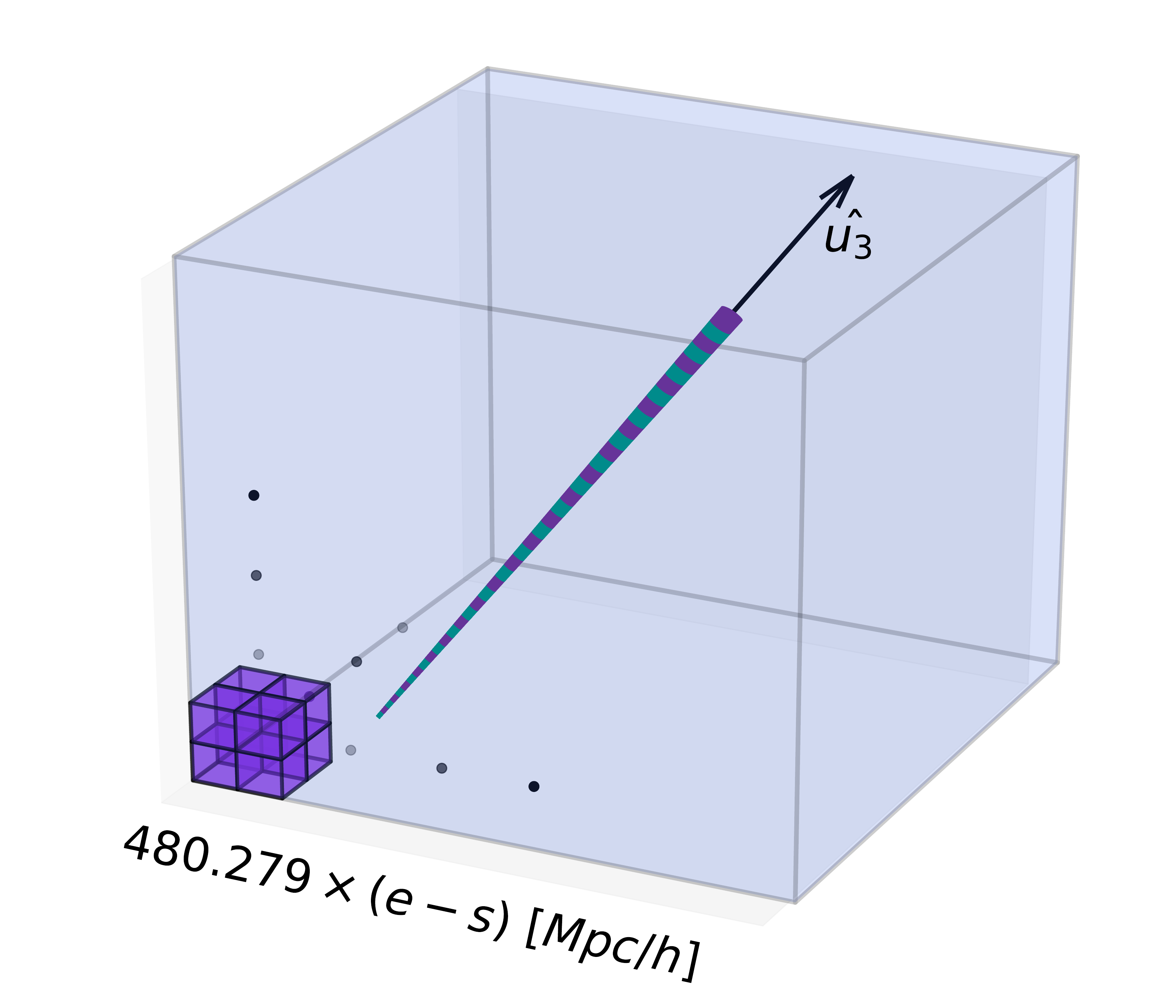}
    \caption{\small The entire lightcone volume contains $(e-s)^3$ complete Millennium volumes and represents a huge volume with side: $(e-s) \times 480.279$ Mpc$/h$. The Figure also shows the lightcone within the created volume. If the snapshot $s_j$ is an odd or even number, then the volume where we select galaxies is represented in purple or cyan, respectively.}
    \label{tot}
\end{figure}

Each Millennium snapshot is a box of comoving side $L=480.279 $ Mpc$/h$, containing the evolutionary stage of the same universe volume at a certain epoch. It is essential to use the information of all available snapshots towards the redshift limit of the lightcone, $z_f$. We extract galaxies from the snapshot $s_j$ which are located between the comoving distance $d_C(z_j)$ and $d_C(z_{j+1})$, where $z_j$ and $z_{j+1}$, represent the redshift of the snapshot $s_j$ and $s_{j+1}$, respectively. Therefore, we need to estimate the position that represents the comoving distance of every snapshot, $\vec{r_{s}}$, considering the zero-point, $\vec{r_o}$.  
Using Equation \ref{eq:s_pos}, we derive the position where the snapshot $s_j$ starts:
\begin{equation}
\vec{r_{s}}(z_j) = d_C(z_j) \hat{u_3} + \vec{r_o}
\label{eq:s_pos} 
\end{equation}
where $d_C(z_{j})$ is the comoving distance to the redshift that corresponds to the snapshot $s_j$, in units of Mpc/$h$. We construct lightcones of $\pi$ deg$^2$, and, consequently, we remove all galaxies outside a radius of $1.0$ deg from the line-of-sight $\hat{u_3}$. 

Due to the finite size of each snapshot, it is necessary to replicate them. The number of replications depends directly on how deep in redshift the lightcone goes. Therefore, we have created an extended volume which corresponds to a box comprised of $(e-s)^3$ successive Millennium snapshot volumes ($e$ and $s$ explained below). In this new space, galaxy positions can take values between $sL \leq x,y,z \leq eL$.    
Volume replication can generate the repetition of certain positions, implying that the same galaxy could appear twice or more in the lightcone, but at different redshifts. Nevertheless, using the values $n=3$ and $m=4$ for the line-of-sight direction\footnote{Remember that the line-of-sight direction $\hat{u_3}$ is given by $\vec{u_3} = (n,m, nm)$.}, $\hat{u_3}$, we can avoid the galaxy replication effect out to $z \sim 5.0$ \citep{kitzbichler07}.

Since the positions of all SAM output galaxies are stored in $x$, $y$ and $z$ coordinates $\leq L$, we need to transform their coordinates to the extended volume; to do this, we use equation \ref{norm_1}:
\begin{equation}
\centering
\vec{r_c} = \vec{r} - L\vec{i}
\label{norm_1}
\end{equation}
where $\vec{r}$ is the galaxy position in the SAM output and $\vec{i}$ is the 3-D replication index; the $\vec{i}$ vector depends on the replicated volume from where the galaxies were extracted and then placed into the lightcone. 
The replication index takes integer values in the range $(s,s,s) \leq (i_x, i_y, i_z) \leq (e,e,e)$, where $\vec{i}=(s,s,s)$ and $\vec{i} = (e,e,e)$ contain the position of the first and last snapshots, respectively. We can then estimate $s$ and $e$ as: 
\begin{equation}
\begin{split}
    s  &= \texttt{round}\left ( \frac{\vec{r_s}(z = 0)\cdotp \hat{u_3}}{L} \right ) -1 \\
    e & = \texttt{round}\left ( \frac{\vec{r_s}(z = z_f)\cdotp \hat{u_3}}{L} \right )
\end{split}
\label{s_value}
\end{equation}
A representation of the lightcone, the extended volumes, and also of the volumes extracted from each snapshot is presented in Figure \ref{tot}. 

\subsection{Redshifts and angular coordinates} \label{sec:spatial_coords}

The comoving distance of each galaxy is: 
\begin{equation}
\centering
d_{C,{\rm gal}} = (\vec{r_c} - \vec{r_o})\cdotp \hat{u_3}
\label{dc_gal}
\end{equation}
where $\vec{r_c}$ is the galaxy position in the extended volume coordinate system.

For simplicity, the galaxies put in the mock are all those satisfying the condition
$d_C(z_{j}) \leq d_{C,{\rm gal}} < d_C(z_{j+1})$, where $z_j$ represents the redshift of the snapshot $s_j$, i.e.,  those galaxies that are at a comoving distance between the comoving distances of two successive snapshots.

Since the comoving distance is measured along the direction $\hat{u_3}$, to estimate the redshift related to this distance (geometric redshift) we assume that all galaxies with a comoving distance between $d_C(z_k) \leq d_{C,{\rm gal}} \leq d_C(z_k) + 30$ kpc are at $z = z_k$, where $z_k$ takes discrete values between $0$ and $z_f$ spaced by the redshift interval equivalent to $30$ kpc at $z_k$. This approach leads to a curvature in the sky positions of the galaxies. Below we estimate angular coordinates taking into account this spherical projection, although this effect is not discernible for small angles ($\lesssim 3$ degrees).

The projection of the position of the galaxies in the $\{ \hat{u_1}$, $\hat{u_2}$, $\hat{u_3}\}$ space give us the transverse and the radial comoving distances. To estimate the right ascension, $\alpha$, and declination, $\delta$, we use the projection of the position of each galaxy in the $\hat{u_1}$ and $\hat{u_2}$ directions. Following \cite{kitzbichler07}, $\alpha$ and $\delta$ are obtained as
\begin{equation}
\begin{aligned}
\alpha = & \arctan \left ( \frac{\vec{r_c}\cdotp\hat{u_1}}{d_{C,{\rm gal}}} \right )\\
\delta = & \arctan \left ( \frac{\vec{r_c}\cdotp\hat{u_2}}{d_{C,{\rm gal}}} \right )
\end{aligned}
\label{a-d}
\end{equation}

The next step is to include the peculiar motions, by adding the radial velocities of the galaxies to their geometric redshifts. To apply this correction, we first compute the position of all galaxies in the lightcone space, $\vec{r_l}$:
\begin{equation}
\vec{r_l} = [\vec{r_c}\cdotp\hat{u_1},\ \vec{r_c}\cdotp\hat{u_2},\ d_{C,{\rm gal}} ]
\label{r_l}
\end{equation}
and, after, we estimate the radial velocity of a galaxy as $v_r = \hat{r_l}\cdotp\vec{v}$, where $\vec{v}$ is the velocity vector in the Millennium simulation at coordinates ($x,y,z$). The ``observed" redshift $z_{\rm obs}$ of each galaxy is then given by
\begin{equation}
z_{\rm obs} = (1+z_{\rm geo}) \left ( 1 + \frac{v_r}{c} \right ) -1
\label{v_pec}
\end{equation}
where $z_{\rm geo}$ is the redshift associated to the comoving galaxy distance, $v_r$ is its radial velocity, and $c$ is the speed of light.

\begin{figure}
\centering
\includegraphics[width=\columnwidth]{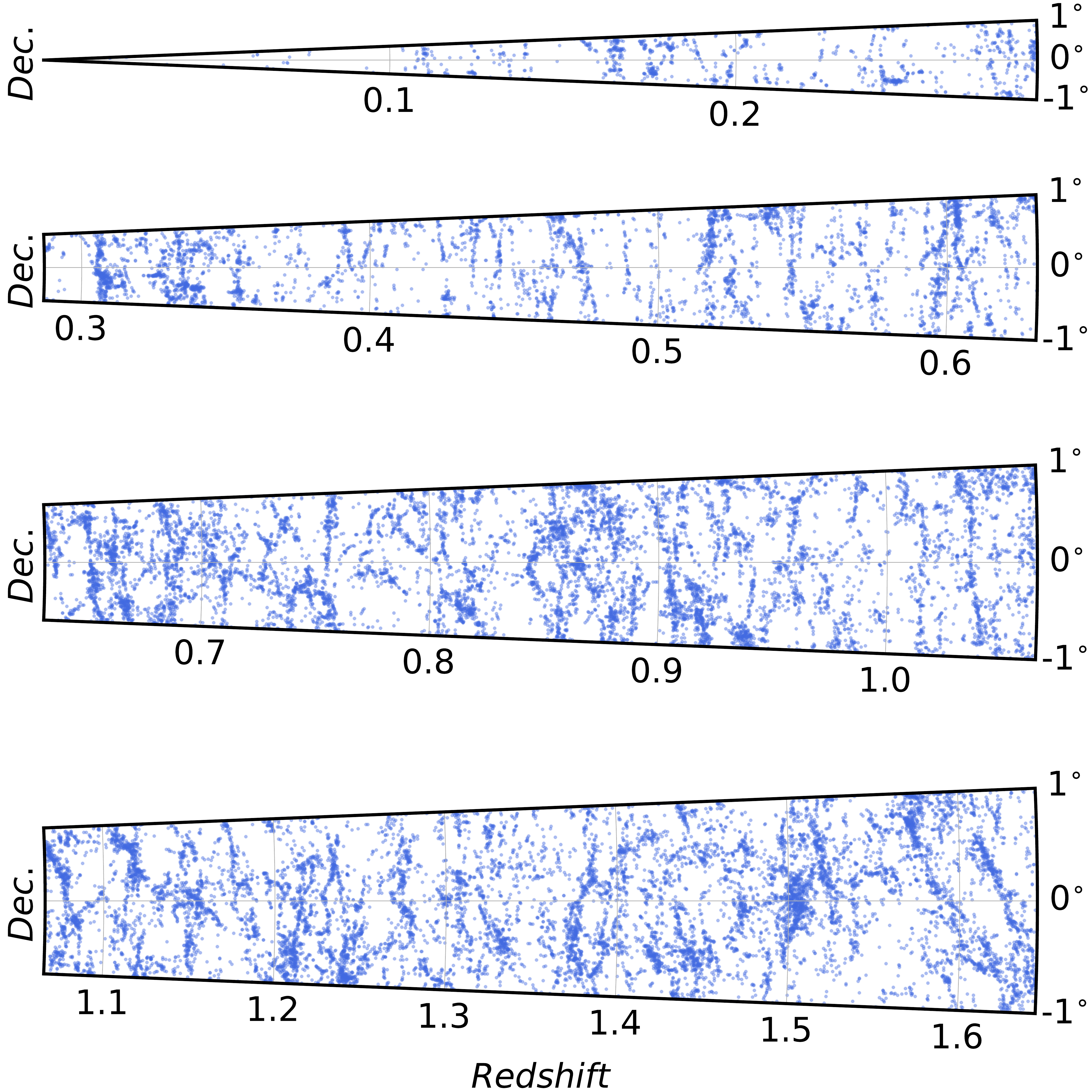}
\caption{\small Lightcone declination against lightcone redshift. We show simulated galaxies brighter than $i < 26.0$ AB mag. The figure presents only galaxies with simulated $\alpha$ between -0.1 and +0.1 degrees.}
\label{lightcone_dist}
\end{figure}

This procedure allows us to obtain the spatial distribution of galaxies in a $\pi$ deg$^2$ lightcone, from $z=0$ to $z=z_f=7.0$. Just for illustration purpose, we show in Figure \ref{lightcone_dist} the declination-redshift distribution between $z=0$ and $z=1.65$ of galaxies brighter than $i = 26.0$ AB mag; we restrict the mock galaxies right ascension to the interval $\alpha = \pm 0.1$ degrees, for better visualization.

\subsection{Galaxy SEDs} \label{sec:galaxy_sed}

By adapting part of \citet{henriques15} SAM code to our lightcone construction script, we estimate spectro-photometric properties of the synthetic galaxies following the \textit{post-processing} prescription of \citet{shamshiri15}, as described below.

As already mentioned, we configure the \texttt{L-GALAXIES} SAM to output the star formation history (SFH) of each galaxy. The SFH comprises two arrays, storing the stellar mass and  metallicity that was produced between two cosmic times for three different baryonic components: disk, bulge, and intra-cluster medium. We assume that disk and bulge SFH bins represent the stellar populations of each galaxy and, with age, metallicity and stellar mass, upon adoption of a stellar synthesis model, we can attribute a spectral energy distribution (SED) for each stellar population in each SFH bin. 

To be consistent with Millennium lightcones, we have used SED templates from  \cite{maraston05} stellar synthesis population models, assuming a \cite{chabrier03} initial mass function. The templates contain $4\times221$ SEDs, with $4$ different metallicities ($\log{(Z/ Z_{\odot})} = -1.35, \ -0.33, \ 0.00, \ 0.35$)
and $221$ different stellar ages (from $\sim 0.2$ Myr to $\sim 20$ Gyr). 
To attribute a SED to each SFH bin, we interpolate the four SEDs that are closest in age and metallicity, and multiply by the stellar mass of the bin.

The dust extinction was modelled following the approach used initially by \cite{delucia07} and, after, by \cite{henriques15}, \cite{shamshiri15} and \cite{clay15}. We present it here just for completeness.
This dust model has two components: the interstellar medium (ISM) and  molecular clouds (MC; actually birth molecular clouds) around newly formed stars. The ISM extinction affects the light from disk  stars, whereas the MC extinction acts on the light from young stellar populations, with ages $\leq 10$ Myr \citep{charlot00}. 
For the first component, the optical depth as a function of the wavelength is:
\begin{equation}
\tau_{\lambda}^{\mathrm{ISM}} = \left ( \frac{A_{\lambda}}{A_v} \right )_{Z_{\odot}}
\left ( \frac{Z_{\rm gas}}{Z_{\odot}} \right )^s (1 + z)^{-1} \times
\left ( \frac{\langle N_H\rangle}{2.1 \times 10^{21} \ \mathrm{atoms \ cm}^{-2}} 
\right )
\label{tau_ism}
\end{equation}
where $\langle N_H \rangle$ is the hydrogen mean column density. This is estimated from  from the SAM output parameters as
\begin{equation}
\langle N_H\rangle = \frac{M_{\rm cold}}{1.4 m_p \pi (a R_{{\rm gas},d})^2} \
\mathrm{atoms \ cm}^{-2} 
\label{n_h}
\end{equation}
where $M_{\rm cold}$ is the mass of the cold gas, $R_{{\rm gas},d}$ the gas disk radius, and $a = 1.68$. Using this value for $a$, $\langle N_H \rangle$ represents the mass-weighted average column density of an exponential disk. The factor $1.4$ accounts for the helium abundance \citep{clay15}.

The $(Z_{\rm gas}/Z_{\odot})$ factor in Equation \ref{n_h} represents the mass fraction of metals in the cold gas, in units of the solar metallicity, $Z_{\odot} = 0.02$. The $s$ parameter depends on the
wavelength: $s=1.35$ for $\lambda < 2000$ \AA , and $s=1.60$ for $\lambda \geq 2000$ \AA~ \citep{guiderdoni87}. The extinction curve for solar metallicity, $(A_{\lambda}/A_v)_{Z_{\odot}}$, is extracted from \cite{mathis83}. 

The second extinction component, MC, affects only young stellar populations, as they are due to the remains of their progenitor molecular clouds. In this case,
\begin{equation}
\tau_{\lambda}^{\mathrm{MC}} = \tau_V ^{\mathrm{ISM}} \left ( \frac{1}{\mu} - 1 \right ) \left ( \frac{\lambda}{5500  \text{ \AA}} \right )^{-0.7}
\label{tau_bc}
\end{equation}
where $\tau_V^{\mathrm{ISM}}$ represents the optical depth of the ISM in the $V$ band ($ \lambda_{eq} \sim 5500  $ \AA ), and $\mu$ is a random Gaussian variable with values between $0.1$ and $1$, with mean $0.3$ and standard deviation $0.2$.

Therefore, we assume that the dust extinction of a galaxy can be written as
\begin{equation}
\begin{split}
A_{\lambda}^{\mathrm{ISM}} &= \left ( \frac{1 - e^{(-\tau_{\lambda}^{\mathrm{ISM}}\sec{\theta})}}{ \tau_{\lambda}^{\mathrm{ISM}}\sec{\theta}} \right ) \\
A_{\lambda}^{\mathrm{MC}} & = (1 - e^{-\tau_{\lambda}^{\mathrm{MC}}})
\end{split}
\label{dusts}
\end{equation}
where $\theta$ represents the inclination of the galaxy.  The inclination cosine is first randomly sampled between $0$ and $1$ and, after, all values smaller than $0.2$ are set to $0.2$ \citep{henriques15}. 

Finally, dust is incorporated into the SEDs of each galaxy by applying the ISM dust factor to the disk total luminosity and the MC factor to the luminosity of young stellar populations in the disk and/or bulge. 
In this work, the SEDs do not include the flux contribution of nebular emission lines.

\subsection{Magnitude Estimations} \label{sec:mags}
Magnitudes were computed from redshifted SEDs following the \textit{post-processing} approach \citep{shamshiri15}. It can be shown that the root-mean-square difference between these magnitudes and those computed during the SAM run time  does not exceed $0.12$ mag for the $u$ band (the worst case for optical bands) for $z=2.0$ galaxies and that this difference decreases for redder filters ($0.02$ mag in the IRAC-4.5$\mu m$ filter) and lower redshifts.

Since we have attributed a SED to each mock galaxy, we can directly obtain apparent magnitudes the considering the filter response function and the galaxy redshift estimated as discussed in Section \ref{sec:galaxy_sed}. The observer-frame flux $S_{\nu}$ is given by
\begin{equation}
    S_{\nu} = (1+z) \frac{L_{(1+z)\nu}}{4\pi d_L(z)^2}
    \label{eq:flux}
\end{equation}
where $L_{(1+z)\nu}$ is the luminosity at the frequency $(1+z)\nu$ for a galaxy at redshift $z$ and luminosity distance $d_L(z)$. 

We estimate apparent magnitudes in the AB system \citep{oke83} as
\begin{equation}
    m_{\rm{AB}} = -2.5\log_{10} \left [ \frac{\int S_{\nu}R(\nu) d\nu}{S_{\rm{o}} \int R(\nu) d \nu} \right ]
    \label{eq:mag_ap}
\end{equation}
where $R(\nu)$ is the filter transmission and $S_{\rm{o}}$ is the zero-point of the AB system, $S_{\rm{o}} = 3631 $ Jy. 

We compute magnitudes for $31$ broad bands from FUV to NIR. Despite the filter transmission of HSC and LSST being similar, we use the filter transmission curves corresponding to each instrument. Notice that we did not include the flux contribution of nebular emission lines or heated dust. Additionally, apparent magnitudes of galaxies with expected prominent emission lines (e.g., Ly$\alpha$, [OII], and H$\alpha$), such as starbursts, are underestimated. Also, we did not incorporate Galactic extinction into the apparent magnitudes, implying that PCcones, when applied to observations, should be used with extinction corrected magnitudes. 

\subsection{IGM Absorption} \label{sec:igm_a}

Magnitudes and colours of high redshift galaxies are critically affected by the absorption of rest-frame UV photons in the intergalactic medium (IGM), mostly those that go through neutral hydrogen clouds, optically thin systems, and resonant scattering of Lyman transitions.  Photons with a wavelength shorter than 1216 \AA \ (Ly$\alpha$) are easily absorbed by neutral hydrogen because they excite or ionize HI atoms, producing a deficit on the observed flux that is more critical for energies  higher than those of the Lyman Limit (912\AA). In consequence, we cannot detect, in the observer-frame, certain objects in the bluer filters. For example, a $z = 3.0$ galaxy will be an $u$-dropout. 

In order to construct a more realistic mock catalog, we adopt the same method as \citet{overzier13} for IGM absorption correction, which we summarize here for completeness. 
\citet{overzier13} correct the mock magnitudes after the filter convolution, as \textit{post-processing}. \texttt{L-GALAXIES} does not bring information on the IGM gas density. However, this quantity can be statistically addressed by an IGM correction based on Monte Carlo simulations, where an effective optical depth is computed from the sum of the contributions of the attenuation sources, using the \citet{inoue08} IGM model. Since the number of absorbers along a certain line-of-sight depends on the redshift, attenuation curves are computed in redshift bins spaced by 0.1, using the \texttt{IGMtransmission} code \citep{harrison11}. This code performs Monte Carlo simulations for 10,000 different line-of-sights and then estimate the effective optical depth. The attenuation curves are applied on a 100 Myr old, continuously star-forming, solar metallicity  SED, modeled using \texttt{starburst99} \citep{leitherer95}. After, the differences between the IGM corrected and the intrinsic magnitudes in all photometric filters are determined. As discussed in \citet{overzier13}, the difference between a  3 Myr old instantaneous low-metallicity starburst galaxy and the current model is less than 0.05 mag. 
\subsection{Protocluster-Lightcones} \label{sec:pcones}

A particularity of our lightcones is that we can place desired structures at specific redshifts. 

\begin{figure}
    \centering
    \includegraphics[width=\columnwidth]{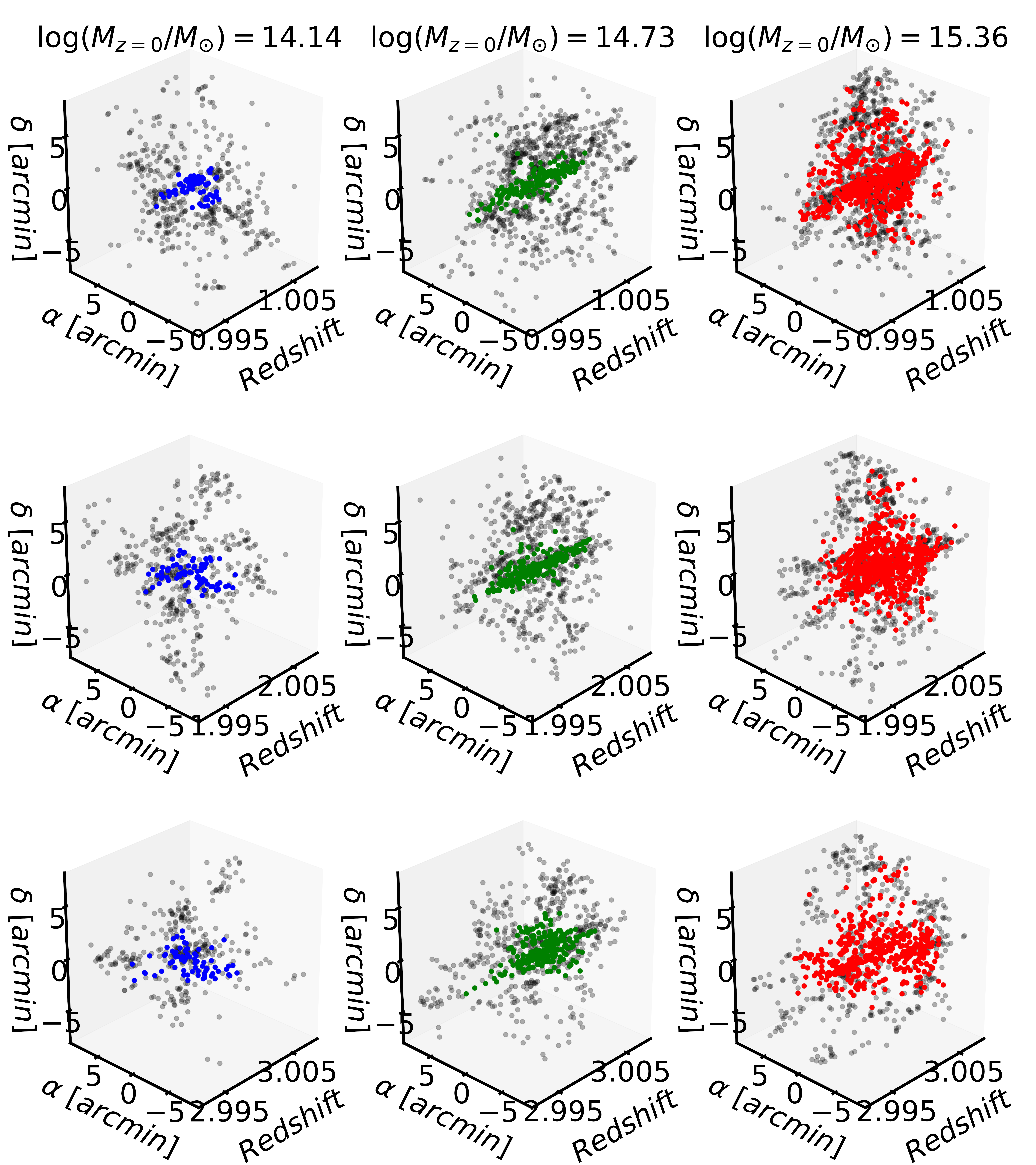}
    \caption{3-D spatial celestial coordinate distribution (right ascension ($\alpha$), declination ($\delta$) and redshift) of three placed (proto)clusters at $z=1.0$, 2.0 and 3.0. coloured dots represents the protocluster members.}
    \label{frame_3d}
\end{figure}

We have chosen randomly 20 Millennium $z=0$ galaxy clusters, and considered the progenitor of these 20 clusters at $z=1.0$, 1.5, 2.0, 2.5, and 3.0. Following the definition of \citet{chiang1}, we have 8 Fornax-type ($M_{z=0} = 1.37-3.00 \times 10^{14} \ M_{\odot}$ ), 6 Virgo-type ($M_{z=0} = 3-10 \times 10^{14} \ M_{\odot}$ ) and 6 Coma-type ($M_{z=0} \geq  10^{15} \ M_{\odot}$ ) protoclusters. We present in Figure \ref{frame_3d} the "observational" 3-D coordinates of 3 of these (proto)clusters at $z=1.0$, $2.0$, and $3.0$, where we highlight the galaxy members of these structures. We select the (proto)cluster members as all galaxies that reside in the dark matter halos that will evolve into the chosen cluster. We present in Table \ref{tab:protoc_inf}  some observational properties of these protoclusters. We estimate the median (proto)cluster galaxy angular distance to the centroid ($\delta \theta_p$), the velocity dispersion ($c\sigma_z/(1 + z)$ where $\sigma_z$ is the standard deviation of the radial velocities of (proto)cluster galaxies), and the number of members with $M_{\star} \geq 10^8$ $M_{\odot}/h$ ($N_{\rm gal}$).  Since these quantities depend on the evolutionary state of the structure, we present them at the five different redshifts.   

This type of lightcone is useful to increase the number of rare structures at specific redshifts, in particular the progenitors of massive clusters (Coma-type). Besides, they can help to track the evolution of structures from an observational point of view, as shown in Table \ref{tab:protoc_inf}.

\begin{table*}
	\centering
	\caption{\small (Proto)cluster descendant mass, median distance from (proto)cluster members to the cluster centroid, galaxy velocity dispersion, and number of (proto)cluster members in the 20 structures at $z=1.0$, 1.5, 2.0, 2.5 and 3.0.}
	\label{tab:protoc_inf}
	\begin{adjustbox}{width=\textwidth}
	\begin{tabular}{c|ccccc|ccccc|ccccc} % four columns, alignment for each
	\hline
	  \multicolumn{1}{c|}{}& \multicolumn{5}{c|}{$\delta \theta_p$ [arcmin]} & \multicolumn{5}{c|}{$c\sigma/(1 + z)$ [km/s]}&\multicolumn{5}{c}{$N_{\rm gal}$} \\
	 		%\hline
       $\log{(M_{z=0}/M_{\odot})}$&$z=1.0$&$z=1.5$&$z=2.0$&$z=2.5$&$z=3.0$&$z=1.0$&$z=1.5$&$z=2.0$&$z=2.5$&$z=3.0$&$z=1.0$&$z=1.5$&$z=2.0$&$z=2.5$&$z=3.0$ \\
         \hline
14.14 & 0.80 & 1.40 & 1.62 & 1.77 & 1.80 & 257.79 & 254.71 & 231.02 & 141.98 & 165.27 & 74 & 74 & 76 & 80 & 78 \\
14.17 & 0.80 & 1.25 & 1.47 & 1.59 & 1.59 & 467.73 & 278.84 & 199.33 & 193.88 & 141.79 & 97 & 101 & 93 & 90 & 88 \\
14.20 & 0.78 & 0.92 & 1.03 & 1.19 & 1.22 & 380.46 & 245.03 & 322.52 & 205.25 & 217.87 & 75& 80 & 95 & 94 & 85 \\
14.23 & 0.76 & 1.08 & 1.17 & 1.11 & 1.06 & 398.82 & 260.19 & 212.49 & 160.65 & 128.92 & 74 & 71 & 77 & 71 & 74 \\
14.27 & 1.89 & 2.22 & 2.49 & 2.34 & 2.18 & 276.48 & 206.40 & 165.45 & 140.33 & 127.16 & 81 & 91 & 95 & 95 & 82 \\
14.31 & 1.15 & 1.78 & 1.55 & 1.77 & 1.71 & 364.60 & 269.56 & 199.26 & 194.50 & 165.96 & 90 & 111 & 111 & 111 & 106 \\
14.36 & 0.83 & 1.06 & 1.34 & 1.53 & 1.74 & 307.77 & 282.04 & 280.35 & 210.81 & 184.77 & 54 & 59 & 57 & 60 & 56 \\
14.42 & 1.52 & 1.60 & 2.19 & 2.24 & 2.24 & 416.96 & 362.18 & 284.29 & 233.02 & 177.78 & 125 & 137 & 141 & 127 & 114 \\
14.49 & 0.92 & 1.09 & 1.13 & 1.25 & 1.44 & 407.09 & 396.81 & 281.18 & 243.42 & 242.17 & 109 & 129 & 119 & 131 & 116 \\
14.58 & 1.41 & 1.21 & 1.37 & 1.38 & 1.62 & 376.47 & 325.81 & 243.41 & 225.44 & 212.62 & 109 & 116 & 122 & 115 & 97 \\
14.73 & 0.89 & 1.18 & 1.41 & 1.64 & 1.77 & 647.35 & 500.49 & 443.68 & 325.70 & 297.03 & 209 & 251 & 262 & 269 & 263 \\
14.73 & 1.80 & 3.22 & 3.11 & 3.05 & 2.82 & 421.83 & 270.74 & 247.66 & 210.62 & 223.00 & 205 & 235 & 239 & 218 & 198 \\
14.82 & 1.36 & 1.64 & 1.83 & 1.98 & 1.89 & 527.48 & 377.53 & 296.57 & 316.65 & 356.97 & 296 & 330 & 322 & 344 & 321 \\
14.94 & 1.62 & 2.39 & 2.42 & 2.68 & 2.66 & 626.69 & 489.45 & 415.85 & 282.60 & 249.00 & 385 & 443 & 438 & 423 & 402 \\
15.03 & 2.69 & 2.52 & 2.56 & 2.61 & 2.62 & 542.45 & 398.31 & 326.22 & 334.36 & 335.15 & 352 & 369 & 365 & 354 & 344 \\
15.05 & 0.92 & 1.39 & 1.95 & 2.29 & 2.38 & 918.87 & 589.27 & 365.95 & 273.45 & 258.58 & 432 & 466 & 462 & 443 & 431 \\
15.07 & 3.18 & 3.85 & 4.05 & 4.04 & 3.81 & 548.90 & 425.99 & 396.73 & 370.91 & 324.81 & 402 & 426 & 423 & 411 & 389 \\
15.16 & 4.16 & 3.76 & 3.45 & 3.47 & 3.35 & 508.83 & 450.26 & 339.91 & 341.90 & 320.92 & 451 & 448 & 442 & 437 & 414 \\
15.26 & 1.98 & 2.12 & 1.91 & 2.08 & 1.97 & 452.64 & 299.46 & 273.02 & 288.02 & 266.29 & 300 & 292 & 291 & 272 & 266 \\
15.36 & 2.78 & 3.56 & 3.47 & 3.54 & 3.56 & 651.56 & 481.89 & 427.88 & 457.91 & 504.98 & 901 & 973 & 960 & 918 & 839 \\

		\hline
	\end{tabular}
	\end{adjustbox}
\end{table*}

\section{Validation Tests} \label{sec:lc_tests}
\begin{figure*}
\centering
\includegraphics[width=\textwidth]{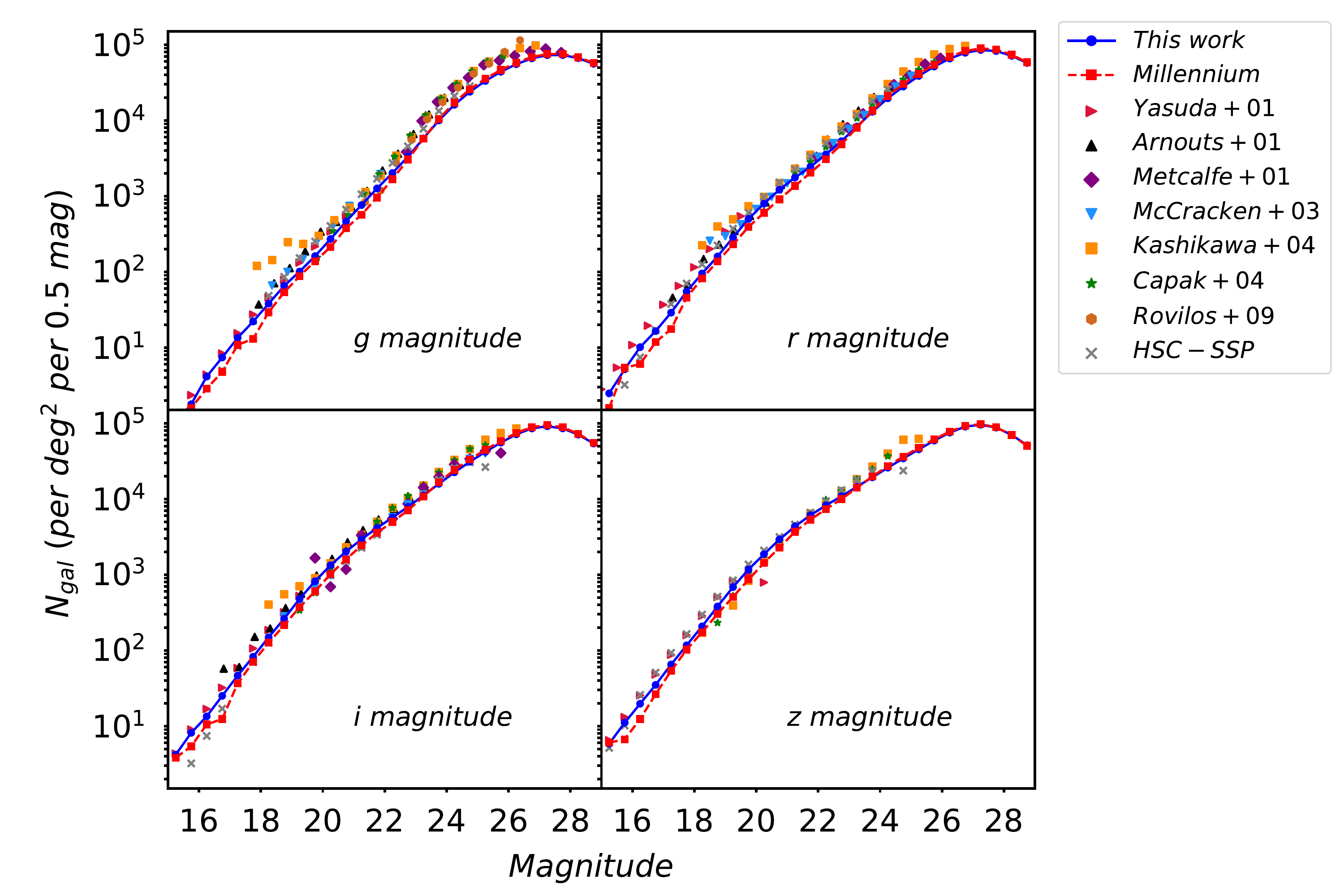}
\caption{\small Number of galaxies per magnitude bin of width $0.5$ per $1$ deg$^2$ of four Sloan filters: $g$ (upper left), $r$ (upper right), $i$ (bottom left) and $z$ (bottom right) bands. Both simulated magnitude distributions (blue solid and red dashed lines for this work and the Millennium lightcone, respectively) are compared with observational data (other symbols) from: \citet{yasuda01}, \citet{arnouts01}, \citet{metcalfe00}, \citet{mccracken03}, \citet{kashikawa04}, \citet{capak04}, \citet{rovilos09} and HSC-SSP data \citep{aihara17}.}
\label{g_count_mag}
\end{figure*}

To test the agreement of the PCcones with observations and other publicly available mock catalogs, we now compare magnitude and colour distributions as a function of the redshift. We also compare our results with those obtained from the lightcones constructed with the \citet{henriques15} SAM, which uses the \texttt{MoMaF} code to obtain observer-frame magnitudes by interpolation.

\subsection{Galaxy Counts}

Figure \ref{g_count_mag} compares galaxy differential number counts as a function of the magnitude in several bands, with observational data from a variety of sources\footnote{data from \url{http://astro.dur.ac.uk/~nm/pubhtml/counts/}}. We compare the predicted magnitude distributions with data from \citet{yasuda01}, \citet{kashikawa04}, \citet{capak04} and HSC-SSP  \citep{aihara17} for the four photometric bands, while for $g$, $r$ and $i$ filters, we also show \citet{arnouts01, metcalfe00} and \citet{mccracken03} data. \citet{rovilos09} galaxy counts are presented for the $g$ band.

Galaxies with stellar masses higher than $10^{8}$ $M_{\odot}/h$ were used to construct predicted galaxy counts for this work and for the Millennium Lightcone and, as expected, both are consistent with each other, since they were computed using the same simulation. The agreement between modeled and observed counts  in Figure \ref{g_count_mag} seems good. However, we can notice that mock counts are systematically lower than the observations. At the faint side, this might be due to stars misclassified as galaxies \citep{capak04}, although the number of faint stars in these fields is expected to be low \citep{kashikawa04}.  In the magnitude interval between 20 and 24.5 mag, the median magnitude correction to match model and observed counts for the four SDSS bands is $\Delta g = -0.36$ mag, $\Delta r = -0.26$ mag, $\Delta i = -0.16$ mag, and $\Delta z = -0.14$ mag. The largest offset is in the $g$-band and might be due to the fact that all galaxy counts in this filter (but \citet{yasuda01} and HSC-SSP data) come from transformations of $B$-band magnitudes. This suggests that these offsets may be due to actual differences in zero points and filter transmissions. The impact of these differences in our analysis is small, and we did not apply any correction to our model magnitudes.

\subsection{Color-Redshift Distributions}

\begin{figure}
\centering
\includegraphics[width=\columnwidth]{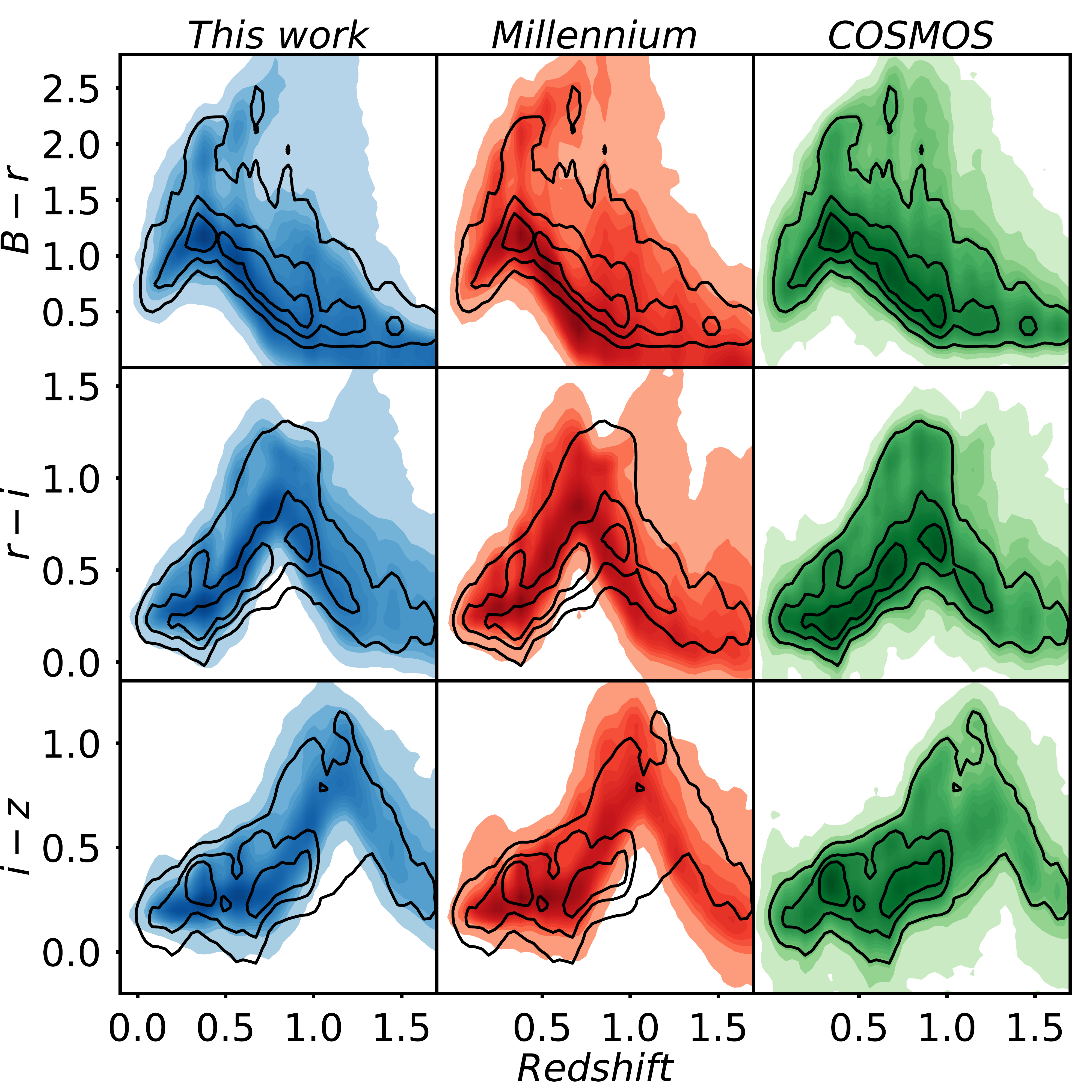}
\caption{\small Colour as a function of redshift for our lightcones (first column), Millennium lightcones (second column), and for the COSMOS2015 data \citep{laigle16} (third column). Each row represents a different colour: $B - r$ (first row), $r-i$ (second row) and $i-z$ (third row). Black contours represent the percentiles 70\%, 90\% and 95\% of the COSMOS2015 distribution (third column).}
\label{color_dist}
\end{figure}

We show in Figure \ref{color_dist} the $B - r$, $r -i$, and $i - z$ colours as a function of the redshift for our mocks, for the Millennium simulation, and for the COSMOS2015 data \citep{laigle16}. The black contours correspond to the 70\%, 90\%, and 95\%  percentiles of the COSMOS distribution. The two main galaxy populations, red and blue galaxies, can be seen in these colour-redshift distributions, both in predicted distributions (this work and the Millennium lightcones) as well as in those obtained from COSMOS data. This figure  shows good qualitative agreement between simulations and observational data, although with some scatter.

\subsection{Color-magnitude Diagram}

\begin{figure}
\centering
\includegraphics[width=\columnwidth]{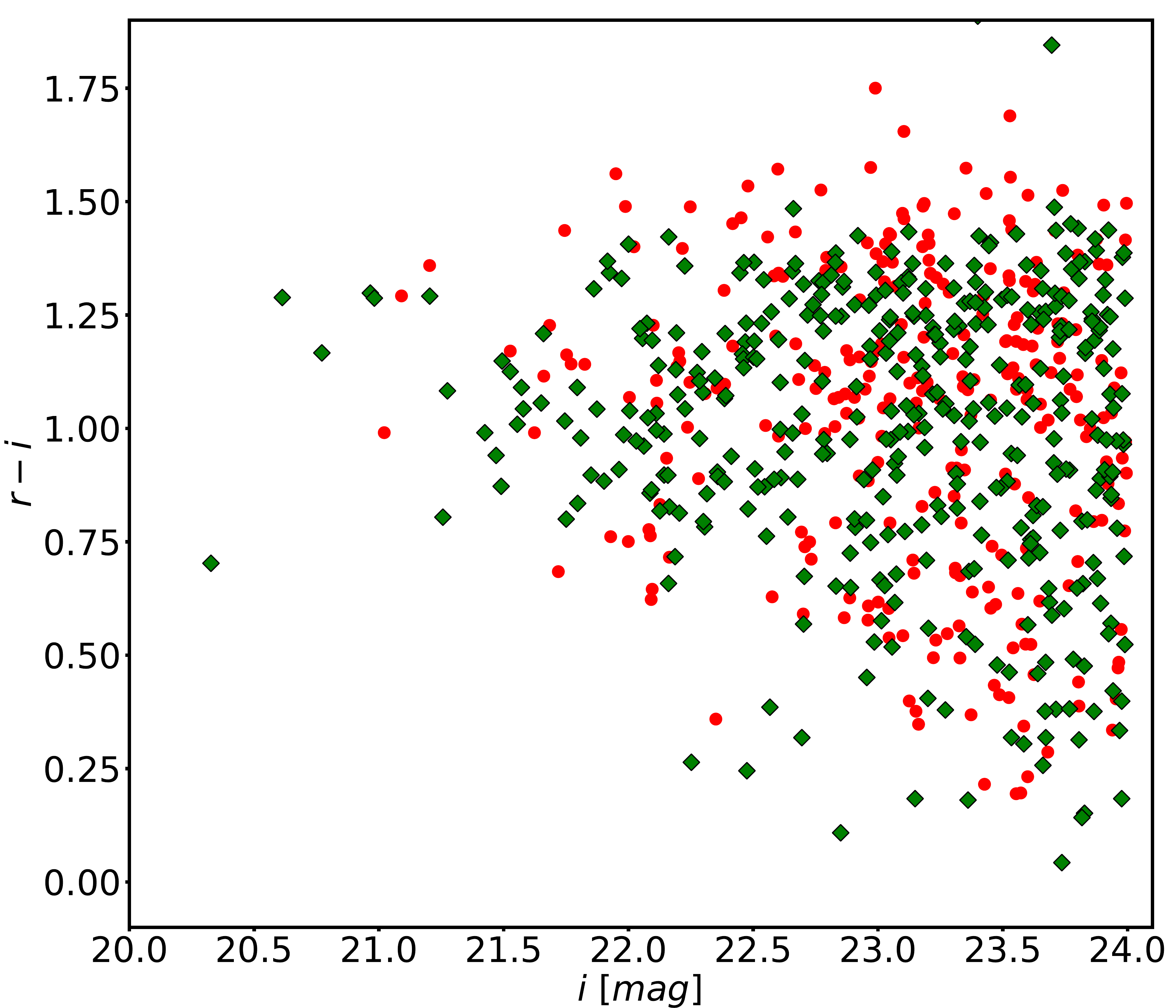}
\caption{\small Colour magnitude diagram for high redshift clusters members between $1.0 \leq z \leq 1.5$. Red dots corresponds to galaxy members from observational data, while green diamonds are mock galaxies.}
\label{color_mag}
\end{figure}
PCcones focus on representing observations of pre-selected structures at a given redshift. For this reason, we compare here the colour-magnitude diagram of the galaxy members of high-z clusters with observational data at $1.0 \leq z \leq 1.5$. For this exercise, we first collect a set of spectroscopic information in the literature, composed by the publicly available catalogs of \citet{tanaka18}\footnote{\citet{tanaka18} catalog is a compilation of public spectroscopic data from: zCOSMOS DR3 \citep{lilly09}, UDSz \citep{bradshaw13, mclure13}, 3D-HST \citep{skelton14, momcheva16}, FMOS-COSMOS \citep{silverman15}, VVDS \citep{lefevre13}, VIPERS PDR1 \citep{Garilli14}, SDSS DR12 \citep{alam15}, SDSS DR14 \citep{paris18}, GAMA DR2 \citep{Liske15}, WiggleZ \citep{drinkwater10}, DEEP2 DR4 \citep{davis03, newman13}, DEEP3 \citep{cooper11, cooper12}, and PRIMUS DR1 \citep{coil11, cool13}.}, VANDELS \citep{mclure18}, and the first data release of GCLASS and GOGREEN surveys \citep{muzzin12, balogh17, balogh2021}. After, we correlate our spectroscopic sample with the photometric catalog of \citet{vanderburg20} by performing a cross-match within 1.0 arcsec. Finally, for simplicity, we define as members all galaxies within a 1.5 cMpc angular radius and $\Delta v \leq 1500$ km/s centred at the position of all GOGREEN/GCLASS cluster candidates. We also apply this criterion for select mock galaxy members. In this case, we used just the PCcones with Coma-type protoclusters at the centre of the mocks. These structures already have masses above $10^{14} \ M_{\odot}$ (high-z clusters), which are the systems analyzed by GOGREEN and GCLASS surveys. 
Figure \ref{color_mag} shows a colour-magnitude diagram of all high-z cluster members of our spectroscopic sample (red dots) and those selected from the mock (green diamonds). This figure presents satisfactory agreement between these two samples for galaxies in dense environments (there is no significant statistical difference between them). Notice that both simulated and observed distributions present evidence of a red-sequence (as a galaxy excess for $(r-i) \gtrsim 1$). 

Figures \ref{color_dist} and \ref{color_mag} present good qualitative agreement between observed and simulated data, what is reassuring, considering that a major motivation for numerical simulations is to produce mock catalogues as realistic as possible, to forecast galaxy survey observations.

\section{Photometric Redshifts} \label{sec:photo_z}
Photometric redshifts are often adopted as a tool to identify structures in the universe, either from targeted observations \citep[e.g.][]{overzier08, hatch17, watson19, strazzullo19}, or in photometric surveys \citep[e.g.][]{omill11, sanchez14, bilicki18, molino19b}. Photometric redshifts are also useful to find galaxy overdensities, and mock catalogues can then be used to quantify the uncertainties of this approach. \citet{chiang1} have examined the impact on simulated protocluster overdensities due to redshift uncertainties, showing that, depending on the galaxy tracer population used, the size of the region, and the redshift accuracy, random regions can present similar overdensities as cluster progenitors at $z=3$.  The back- and foreground contamination is often large in photometrically selected overdense regions, as can be seen with a spectroscopic follow-up \citep[e.g.][]{dey16}. 

We now readdress the impact on  overdensity estimations of uncertainties in photometric redshifts with our PCcones. For this, we will apply to the mocks the same constraints of observational surveys. We will emulate three optical photometric surveys: the Deep Canada-France Hawaii Telescope Legacy Survey (CFHTLS; \url{https://www.cfht.hawaii.edu/Science/CFHTLS/}), which observed 3.2 deg$^2$ of the sky in 5 photometric bands; the ongoing Hyper Suprime-Cam Subaru Strategic Program (HSC-SSP) \citep{aihara17}, whose wide-layer survey will cover 1,400 deg$^2$, whith more than 300 deg$^2$ already observed; and, finally, the future Vera Rubin Observatory Legacy Survey of Space and Time (LSST) \citep{ivezic08}, which will provide deep photometric information in 6 bands over $\sim$20,000 deg$^2$ of the Southern sky after operation over a 10-year period. 

We adopt for the photometric redshift estimation the   \texttt{Le Phare} software \citep{arnouts02, ilbert06}, using a configuration similar to that adopted by \citet{ilbert09}, and as well as by \citet{ilbert13}, \citet{laigle16} and \citet{laigle19}. The software fits SEDs from a set of 31 templates, including spiral and elliptical galaxies from \citet{polletta07} (a total of 19) and also 12 young blue star-forming galaxies modeled with \citet{bruzual03} stellar population SEDs. We added dust extinction as a free parameter ($E( B - V)  < 0.5$), and considered different extinction laws: \citet{calzetti00}, \citet{prevot84}, and the Calzetti laws including a bump at 2175 \AA \  \citep{Fitzpatrick86}. Following \citet{laigle19}, who performed photo-z estimates for simulated galaxies without including emission lines, we did not add this flux contribution in the templates. 

We have tested the impact of Le Phare error adjustments on our photo-z estimations and, similar to \citet{laigle19}, we did not find significant differences. Therefore, we did not take into account Le Phare systematic errors in the magnitudes (through the ERR\_SCALE parameter). 

To quantify the accuracy of photometric redshifts, we estimate the normalized median absolute deviation, $\sigma_{\rm{NMAD}}$, defined as in \citet{brammer08}, \citet{molino14}, and \citet{molino19}. Also, we quantify the bias, $b$, and the outlier fraction, $f_{\rm{outliers}}$, following \citet{ilbert06}, \citet{ilbert09}, and \citet{tanaka18}. These quantities are defined as
\begin{equation}
\raggedright
    \begin{split}
    \sigma_{\rm{NMAD}} &= 1.48 \times median \left ( \frac{| \delta z  - median(\delta z) |}{1+z_s}  \right )  \\
    f_{\rm{outliers}} &= \frac{N[|\delta_z|/(1 + z_r) > 0.15]}{N_{total}} \\
    b &= \langle \delta z \rangle 
\end{split}
\label{eq:photo-zs}
\end{equation}
where $\delta z = z_p - z_r$, with $z_p$ denoting the estimated photometric redshift and $z_r$ the reference redshift of the galaxies in the lightcone.

\subsection{Generating realistic observed magnitudes} \label{sec:o-likemags}

Observed galaxy magnitudes are affected by many factors, such as the exposure time, sky brightness, quantum efficiency of the detector, the point spread function, blending, reduction artifacts, etc, that can introduce systematic and random errors in the measurements. In order to emulate observational magnitudes in our mocks, we have implemented the same technique as \citet{graham18}, where we assume random errors in the mock magnitudes by using the analytic expression presented in \citet{ivezic19}:

\begin{equation}
    \sigma_{rand}^2 = (0.04 - \gamma)x + \gamma x^2 \ (mag^2),  
    \label{eq:obsmags}
\end{equation}
where $x \equiv 10^{0.4(m - m_5)}$, $m_5$ corresponds to the $5\sigma$ magnitude limit of the observational survey to be emulated, and $\gamma$ depends on sky brightness, signal-to-noise and photometric filter. 
We set $\gamma$ to the same values presented in Table 2 of \citet{ivezic19} for optical magnitudes. The values of $m_5$ and $\gamma$ that we adopt to simulate the deep survey of CFHTLS, the wide survey of HSC-SSP, and LSST, are listed in Table \ref{tab:ground}. We assume a band-dependent $\gamma$ that is the same for the three surveys. Notice that this assumption is acceptable in this case because our photo-z accuracy is consistent with observational results for these surveys (see subsection \ref{sec:photo_zs}). 

\begin{table}
	\centering
	\caption{ \small List of values used for observational like magnitudes estimations for ground based surveys (the deep survey of CFHTLS, the wide survey of HSC-SSP and LSST) for each filter, where $m_5$ represents the $5\sigma$ magnitude limit and $\gamma$ a image quality parameter.}
	\label{tab:ground}
	\begin{tabular}{ccccc} % four columns, alignment for each
		\hline
		Filter & $m_5$ (CFHT)& $m_5$ (HSC)& $m_5$ (LSST)& $\gamma$ \\
		\hline
		u & 26.3 & -  & 26.1 & 0.038\\
		g & 26.0 & 26.8 & 27.4 & 0.039\\
		r & 25.6 & 26.4 & 27.5 & 0.039\\
		i & 25.4 & 26.2 & 27.0 & 0.039\\
		z & 25.0 & 25.4 & 26.1 & 0.039\\		
		y & - & 24.7 & 24.9 & 0.039\\
		\hline
	\end{tabular}
\end{table}

The total magnitude error is obtained as: $\sigma_{mag} = \sqrt{\sigma_{sys}^2 + \sigma_{rand}^2}$, where we have assumed a systematic error of $\sigma_{sys} = 0.005$, following \citet{ivezic19}. If we adopt twice this value, the photometric redshift estimates do not change significantly ($\Delta \sigma_{\rm NMAD < 0.001}$). Notice that real surveys are plagued by additional effects that affect their photometric measurements, such as bright foreground stars, satellite trails, reduction artifacts, among others, which are not considered here. Therefore, implicitly, we are assuming a best-case photometric scenario.

\begin{figure}
\centering
\includegraphics[width=\columnwidth]{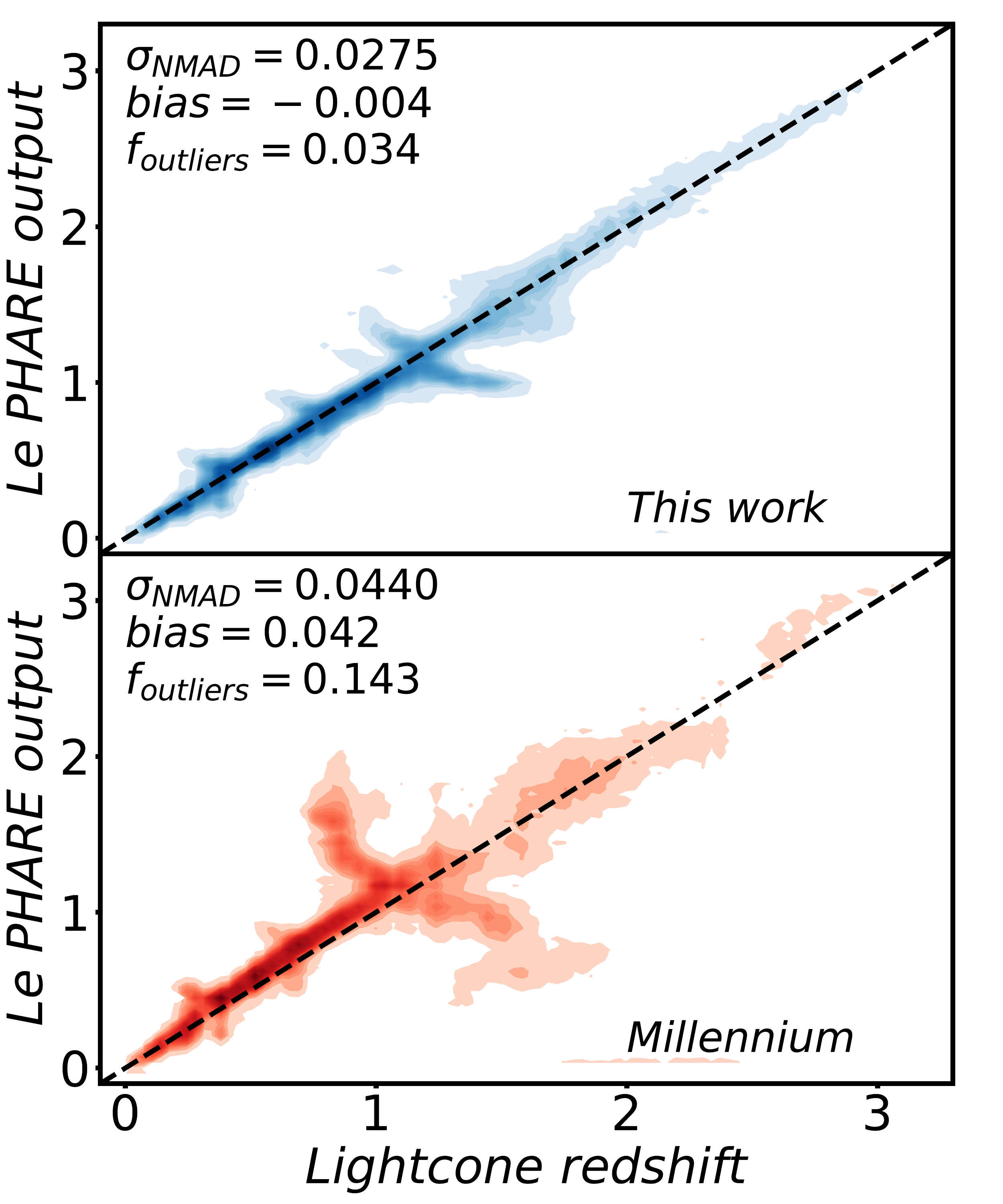}
\caption{\small Photometric redshift estimation using  \texttt{Le Phare} versus the lightcone redshift for this work (blue contours, first panel) and the Millennium Lightcone (red contours, second panel). Photo-zs were estimated from the photometric information of the deep CFHTLS.} 
\label{photo-z}
\end{figure}

\subsubsection{Non-detected sources}

As we have mentioned in Section \ref{sec:igm_a}, the Lyman Break may lead high redshift galaxies not being detected in all photometric filters due to IGM absorption along the line of sight. Also, the depth of each survey plays an important role in the sources that we can observe. Since mock catalogs are limited just by the resolution of the simulation, and this limit allows us to have a complete sample at fainter magnitudes compared with the limits of the surveys discussed here, we need to allow for non-detected sources in some photometric bands. Therefore, we define as a non-detected source all galaxies with signal to noise smaller or equal to 1 for each of the three surveys, and we set this value as an upper limit in the photometric redshift estimation with \texttt{Le Phare}.    
\begin{figure*}
\includegraphics[width=\textwidth]{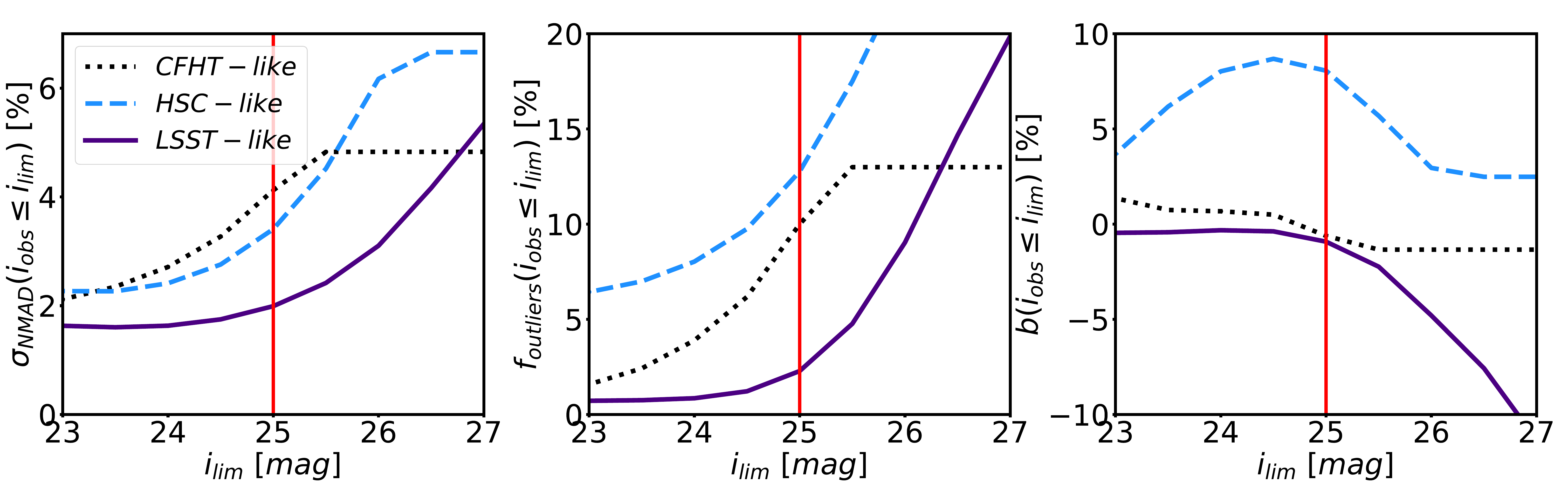}
\caption{\small Distribution of the normalized median absolute deviation ($\sigma_{\rm NMAD}$, first panel), outlier fraction ($f_{\rm outlier}$, second panel), and $bias$ ($b$, third panel) as a function of magnitudes for the CFHT-, HSC-, and LSST-like emulations. Red vertical line indicates the imposed magnitude limit for the full mock sample.}
\label{fig:metrics}
\end{figure*}

\subsection{Comparison with Millennium photometric redshifts} \label{sec:photo_zm}

In this section we compare photometric redshift estimates obtained for PCcones and one (number twenty-two) of the twenty-four the Millennium lightcones. We simulate observations of the deep Canada France Hawaii Telescope Legacy survey using both lightcones. The values of $m_5$ and $\gamma$ for each filter are listed on Table \ref{tab:ground}.  The photo-z estimation performed with \texttt{Le Phare} are presented in Figure \ref{photo-z}.  

Comparing their performance with our three metrics, we obtain  $\sigma_{\rm{NMAD}} = 0.044$, $b = 0.042$ and $f_{\rm{outliers}} = 0.143$ for the Millennium lightcone, while for the photo-z estimation over our lightcones, we achieve $\sigma_{\rm{NMAD}} = 0.028$, $b = -0.004$ and $f_{\rm{outliers}} = 0.034$. 

These results and Figure \ref{photo-z} indicate that photometric redshifts using our lightcones are more reliable than those with the usual Millennium lightcones. Compared to ours, the photo-z estimation using the Millennium apparent magnitudes presents a clear bias for $z \lesssim 1$, where the linear correlation is offset with respect to the equal values line. Additionally, the bias and the outlier fraction obtained with the conventional Millennium simulation are larger than ours.

The unique main difference between both lightcones is the method to estimate apparent magnitudes, in particular, the observer-frame magnitudes. We computed magnitudes in \textit{post-processing}, following \citet{shamshiri15}, attributing a SED to each galaxy, while those of the Millennium Lightcones come from interpolations using  MoMaF \citep{blaizot05}.
To confirm that the differences in photo-z estimates are indeed caused by this, we performed a more detailed comparison. We re-calculated the apparent magnitudes for the galaxies in the Millennium Lightcone at $z \leq 3.0$ and then we estimated again their photometric redshift (see Appendix \ref{sec:comparison}). Our results present the same trend as those we find in Figure \ref{photo-z}. Also, we obtain similar values of $\sigma_{\rm{NMAD}}$, bias, and outlier fraction. We show the analog to Figure \ref{photo-z} in Figure \ref{fig:comparison_mill}. 

We conclude that photometric redshifts using our lightcones are more reliable because our method to obtain observer-frame magnitudes implicitly change the shape of the galaxy SED caused by the spectral deviation, while the interpolated ones correct by the systematic wavelength shift \citep{merson13}. In addition, \texttt{Le Phare} fits real SEDs and, therefore, the differences in the photo-zs estimates could come from the fitting of a real SEDs to an interpolated colour.

\begin{figure*}
\includegraphics[width=\textwidth]{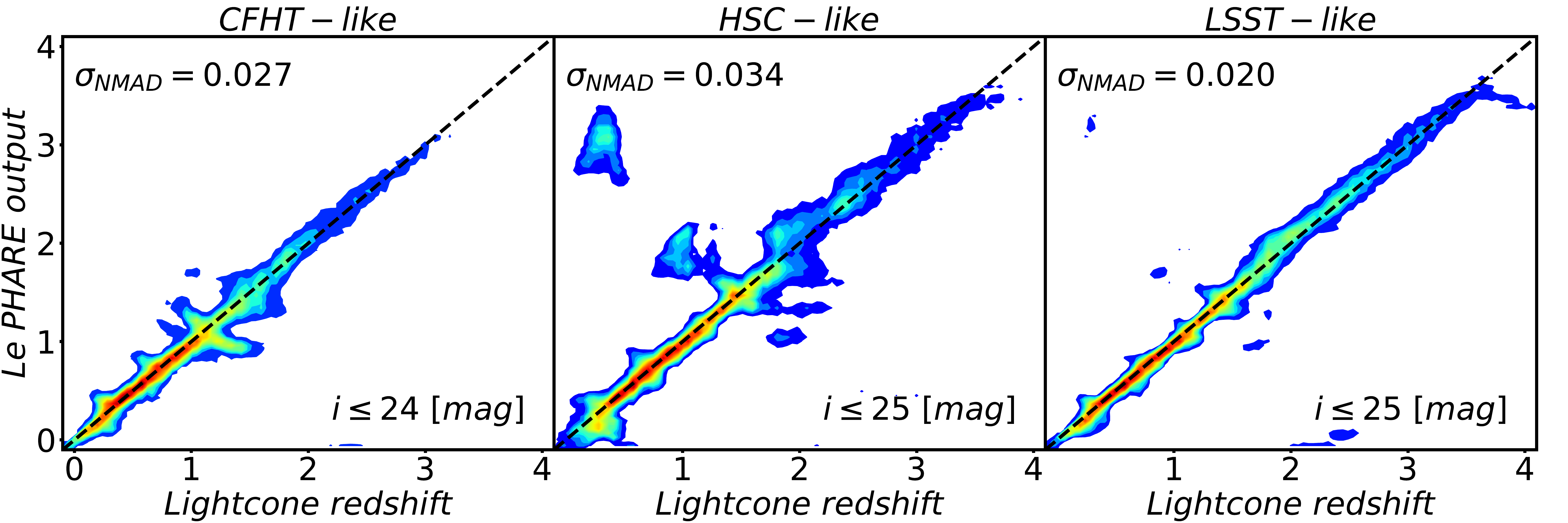}
\caption{\small Photometric redshift estimation using \texttt{Le Phare } for the deep CFHTLS (left panel), the wide layer of HSC-SSP (middle panel), and LSST (right panel) survey emulations.}
\label{photo-z_surveys}
\end{figure*}
\subsection{Photometric Redshifts for Simulated Surveys}\label{sec:photo_zs}

We now use our lightcones to obtain photometric redshift estimates for some actual photometric surveys. In order to mimic observations of the deep CFHT legacy survey, the wide layer of the Hyper Suprime-Cam Strategic Survey Program (HSC-SSP), and LSST, we assume the $5\sigma$ magnitude limits listed in Table \ref{tab:ground} in the photometric bands available for each of these surveys. 

Firstly, we impose a magnitude cut in the $i-$band equals to its $5\sigma$ limit to analyze the photo-z accuracy with our three metrics: $\sigma_{\rm NMAD}$, $f_{\rm outlier}$ and $b$ (see Equation \ref{eq:photo-zs}) for just one mock.  Figure \ref{fig:metrics} show the dependence of these parameters as a function of the magnitude in the $i-$band. This figure recovers an important point in photometric redshift estimation, namely, the improvement in photo-z accuracy ($\sigma_{{\rm NMAD}}$) by increasing the number of bands and by decreasing photometric errors. The outlier fraction and the bias are worst for HSC-SSP. This is due to the lack of the $u$-band, generating confusion between Lyman Break Galaxies at $z\sim 3.0$ and low-z quiescent galaxies. 

Additionally, all the analyzed parameters reach catastrophic values for fainter magnitudes. For example, the bias parameter increases (in absolute value) for LSST at $i > 25.0$ mag, while the outlier fraction and $\sigma_{\rm NMAD}$ also have a fast increase for HSC-SSP and LSST at the same range.  For this reason, we have estimated photometric redshifts for the whole PCcones sample, but constrained to objects brighter than $i = 25.0$ mag for HSC-SSP and LSST mock surveys. In the case of the CFHTLS emulation, we imposed a magnitude cut at $i = 24.0$ mag, the same as \citet{wen11}, who produced a cluster sample that we will emulate here (see Section \ref{sec:cfht-like}). The $i$ band, which is used to make these magnitude cuts, corresponds to the rest-frame far-UV at $z=3.0$. Adopting $L^{\ast}_{\rm FUV}$ from \citet{hathi10}, and considering no k-correction, we obtain that these limits are equivalent to $\gtrsim 0.6L^{\ast}_{\rm FUV}$ for the HSC/LSST-like mocks, and to $\sim1.5L^{\ast}_{\rm FUV}$ for the CFHT-like mock.

We have obtained  $\sigma_{\rm{NMAD}} = 0.027$, $0.033$, and  $0.020$ for CFHTLS, HSC-SSP and LSST-like observations, respectively. Since our definition of accuracy differs from those reported in papers related to these surveys, below we use the normalized redshift dispersion, $\sigma_z /(1+z)$, as adopted in these other studies. With this metric \citet{ilbert06} obtained a redshift accuracy of about $0.029$ for the deep CFHTLS, for galaxies with $i < 24.0$, while we obtained $\sigma_z /(1+z)= 0.027$. Also, \citet{graham18} predicted a photo-z accuracy about $0.017$ for the $10$ years of LSST for galaxies with $i < 25.0$. After applying this same magnitude limit, we achieve $0.020$. For the HSC-like sample, we obtained $0.037$, while \citet{tanaka18} obtained an accuracy of about $0.050$. These results show that our procedure gives photometric redshifts with accuracy comparable to those inferred from the studies mentioned above. 

\citet{ilbert06} obtained an  outlier fraction, $f_{\rm outlier}$ for the Deep CFHTLS of $3.8\%$ at $i < 24.0$ mag, while we achieved  $3.9\%$ with the same magnitude restriction. For the wide-layer of HSC-SSP, \citet{tanaka18} reported that $\sim 15$ \% of the galaxies present catastrophic redshifts, whereas our measurements indicate  $13\%$.  Finally, for the LSST case, \citet{graham18} computed this fraction more strictly than the previously mentioned reports. They considered as outliers the estimates that differ by a factor 0.06 instead of 0.15 in the second row of Equation \ref{eq:photo-zs}. Using the same definition of \citet{graham18}, who found an outlier fraction of 4\%, we predict for the 10-year forecast of LSST a fraction of $\sim 3\%$. Notice that for HSC-SSP and LSST, the applied magnitude limit in this work and the others is $i = 25.0$ mag.
\begin{figure*}
\centering
\includegraphics[width=\textwidth]{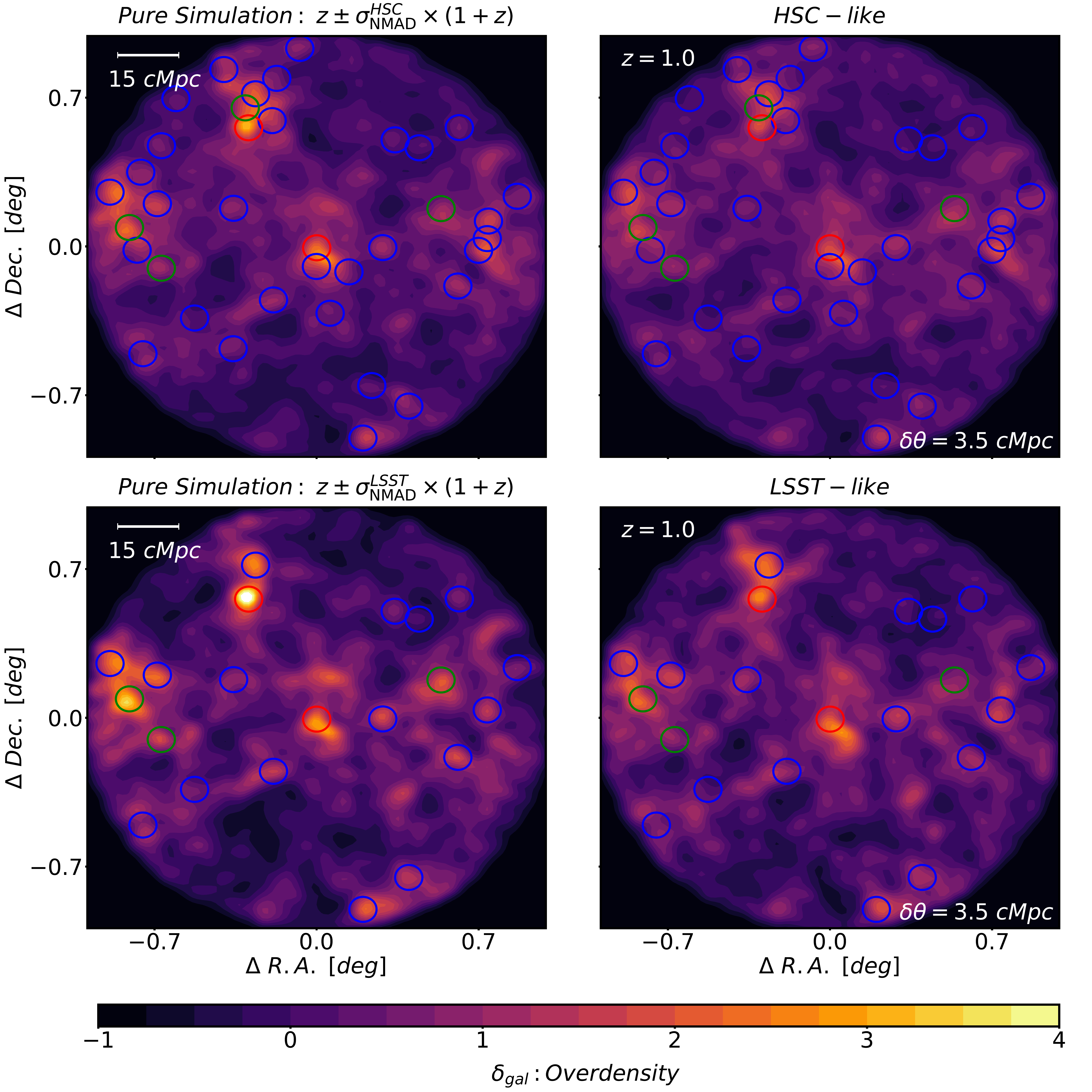}
\caption{\small Example of a surface density map obtained from a PCcone where a $\log{(M_{z=0}/M_{\odot})} = 15.26$ cluster progenitor is placed at the centre of the lightcone at redshift $z_p = 1.0$. We present the density map for the \textit{Pure Simulation} sample (left column) and the \textit{Observational-like} samples (right column). HSC-like simulations are at the top, whereas LSST-like ones are at the bottom  both assuming a magnitude limit $i \leq 25.0$ mag. The colour circles, with radius $R_e$, mark protocluster regions, where the red, green and blue ones indicate Coma, Virgo and Fornax type protoclusters, respectively . Notice that there are several other protoclusters in the slab, besides that put at its centre.}
\label{overdensity_panels}
\end{figure*}

\section{Results} \label{sec:overdensities}

\begin{figure*}
    \centering
    \includegraphics[width=\textwidth]{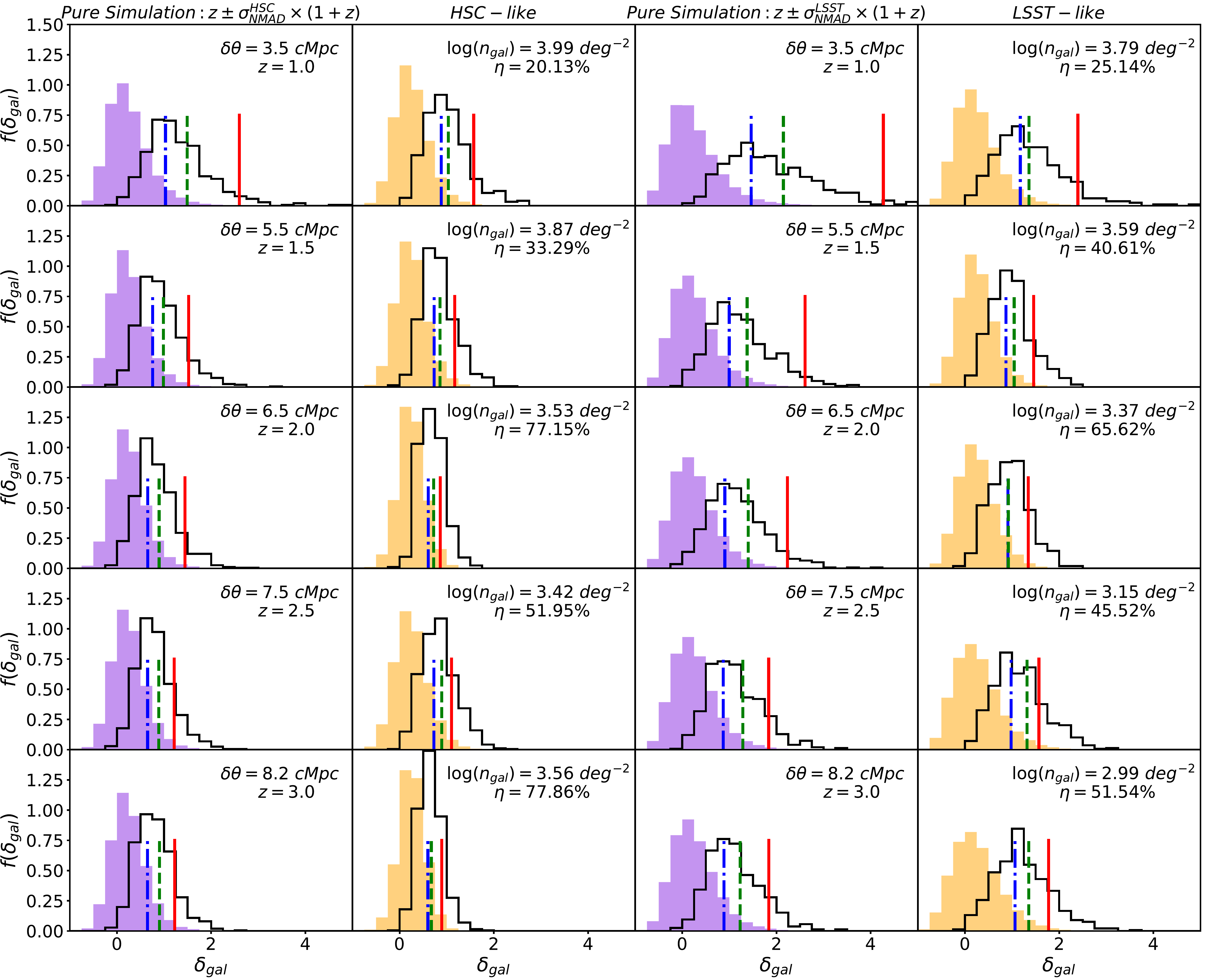}
    \caption{\small Probability density function, $f(\delta_{\rm gal})$, of density contrast, $\delta_{\rm gal}$, of all pixels in the maps in the $\pi$ deg$^2$ field (filled histogram) and associated to protocluster regions (histogram with solid lines) at $z=1.0$, $1.5$, $2.0$, $2.5$ and $3.0$ (from top to bottom) obtained from the \textit{Pure Simulation} (first and third columns), \textit{HSC-like} (second column) and \textit{LSST-like} (fourth column) samples. The median overdensity of Fornax-type (blue dash-dot line), Virgo-type (green dashed line) and Coma-type (red solid line) progenitors is also presented.}
    \label{hist_tot}
\end{figure*}

Measuring the excess of galaxies in a given region of the universe with respect to the mean distribution is one of the most common techniques to look for cluster/protocluster candidates. In this section we explore overdensity measurements taking into account observational constraints on our mocks, assuming a best-case observational scenario, such as described in Section \ref{sec:o-likemags}.  

We shall divide our discussion into two parts: in the first one, we will perform a forecast of structure detection in specific redshift ranges ($z = 1.0,1.5,2.0,2.5$ and $3.0$), emulating the wide layer of the HSC-SSP and the future LSST survey. We compare these mock overdensity distributions, measured in redshift intervals (see below), to those obtained from mocks without magnitude constraints. Also, we estimate the probability of a region with overdensity $\delta_{\rm gal}$ be a protocluster. 

In the second sub-section, we compare the cluster distribution for a CFHTLS mock  with actual observational data in the redshift range $0.12 < z < 1.70$, as an additional test of using our lightcones to predict cluster detections over a wide redshift range.  

\subsection{Structure Detection Forecast}
In this section, we describe our method to detect protoclusters at different redshifts with density contrast maps. We use this information to compute the probabilities of a certain region be a real structure. Also, we analyze the detectability of the protoclusters placed in the central line-of-sight of the PCcones. 
 
\subsubsection{Mock samples}
We perform our analysis for two main different mock samples: the first one (which we call \textit{Pure Simulation}) contains all simulated galaxies (i.e., without any magnitude constraints) within a $\Delta z$ according to the photometric redshift accuracy achieved for each survey, centred at the redshift where we placed the protocluster. Our motivation with this sample is to compare overdensity measurements in mocks without any additional magnitude limits and using redshift from simulations, to estimations from the photo-z surveys emulations. We estimate \textit{Pure Simulation} overdensities within $\Delta z = \sigma_{{\rm NMAD}}^{HSC} \times (1+z) = 0.034 \times (1+z)$  and $\Delta z = \sigma_{{\rm NMAD}}^{LSST} \times (1+z)= 0.020 \times (1+z)$ for HSC and LSST, respectively.

The second type of samples, which we call \textit{Observational-like} samples, is composed by the two mock observational surveys, emulating observations of the HSC-SSP wide layer (hereafter, \textit{HSC-like}) and the $10$-year LSST survey forecast (hereafter, \textit{LSST-like}). 
We have done this by adding noise to the mock magnitudes. As explained in Section \ref{sec:o-likemags}, this depends on the $5\sigma$ magnitude limits of the surveys, listed in Table \ref{tab:ground}. For both surveys, we considered all galaxies with $i \leq 25.0$ mag. We also use the photometric redshifts we have estimated before, which have a mean accuracy of $\sigma_{{\rm NMAD}} = 0.034$ and 0.020 for the \textit{HSC-like} and \textit{LSST-like} samples, respectively, up to the applied magnitude limit.

\subsubsection{Overdensity Estimation}
We estimate overdensities in redshift slabs by selecting galaxies in $z_p \pm \Delta z$,  where $z_p$ is the proto-cluster redshift (1.0, 1.5, 2.0, 2.5, or 3.0)  and $\Delta z$ = $ \sigma_{{\rm NMAD}} \times (1+z)$, similar to \citet{chiang14}, who adopted $\Delta z = 0.0125 \times (1 + z)$, according to the photometric redshift accuracy of the COSMOS/ultraVISTA survey. Within each slab, the density field is computed using the Gaussian Kernel Density Estimator (KDE) of the \texttt{scipy} Python package, version 1.3.2. The kernel bandwidth is set as a function of redshift as $\delta \theta(z) = R_e(z)$, where $R_e(z)$, following \citet{chiang1}, is the effective radius of Coma-type protoclusters. It takes values: $ R_e(z)= 3.5, 5.5, 6.5, 7.5$ and $8.2$ cMpc at $z = 1.0, 1.5, 2.0, 2.5$ and $3.0$, respectively. We will briefly discuss the impact of this bandwidth choice at the end of subsection \ref{sec:pc_probs}.

If $\Sigma$ is the Gaussian kernel estimator output for a given celestial coordinate in a redshift slab, the density contrast at this point, $\delta_{\rm gal}$, is given by:
\begin{equation}
    \delta_{\rm gal} = \frac{\Sigma - \langle \Sigma \rangle}{\langle \Sigma \rangle}
\end{equation}
where $\langle \Sigma \rangle$ denotes the mean of $\Sigma$ inside the slab. Examples of surface density maps are presented in Figure \ref{overdensity_panels} for the two mocks discussed in this section, where the progenitor of a $z=0$ galaxy cluster of $M_{z=0} = 2.28\times 10^{15} M_{\odot}$ has been placed at redshift $z=1.0$. The density contrast maps are computed using a $200\times200$ pixels grid that covers $-1.0 < \Delta RA < 1.0$ [deg] and $-1.0 < \Delta Dec < 1.0$ [deg]

\subsubsection{Pure Simulation Overdensities}
In this section, we analyze the \textit{Pure Simulation} sample, which considers all simulated galaxies in a certain volume, without any photometric constraints. We look for the (proto)clusters in redshift slabs within the lightcone field of view.

In order to know how many protoclusters are in the $100$ lightcones at $z_p \pm \Delta z$, we select all galaxies that reside in dark matter halos that will evolve into a galaxy cluster at $z=0$ or before. Then, we obtain the protocluster positions (right ascension, declination, and redshift) as the median of the distribution of their galaxies. Notice that the volume within a slab in our $\pi$ deg$^2$ mocks can vary with the redshift, but the variation is small in the redshift range discussed here.

In some cases, we find clusters with less than $\sim 10$ galaxies in the redshift interval. This is due to border effects (in area and/or in depth) and may produce an underestimation of the \textit{Pure simulation} overdensities in regions occupied by protoclusters. To avoid it, we will analyze the overdensities of all protoclusters with median redshifts within the redshift slab and angular coordinates inside a radius of $1{\rm deg}-\delta \theta(z)$ from the centre of lightcone. Remember that $\delta \theta(z)$ is the kernel bandwidth, defined as the effective radius of Coma-type protoclusters at redshift $z$.

Figure \ref{hist_tot} shows the density contrast distributions of the field at each redshift slab (filled distributions) that represents the $\delta_{\rm gal}$ of all pixels of the $200\times200$ grid inside a radius of $1$ deg from the centre of the lightcone. In this figure, we present both \textit{Pure Simulation} samples in the first and the third columns, for the two surveys being emulated in this section. It is also shown the density where protocluster galaxies reside. 

\subsubsection{Overdensities for emulated surveys}

Here, we use the \textit{Observational-like} samples, which represent the data sets constrained by the $i \leq 25$ mag magnitude limit for the wide layer of the \textit{HSC-SSP} and \textit{LSST} surveys. The corresponding photometric redshift estimations obtained using \texttt{Le Phare} are shown in Figure \ref{photo-z_surveys}. 

\texttt{Le Phare} output gives the most likely redshift, $z_{\rm phot}$. We have done overdensity maps (see Figure \ref{overdensity_panels}) from the distribution of galaxies within $|z_{\rm phot} - z_p| \leq \sigma_{\rm{NMAD}} \times (1+z)$, where  $\sigma_{\rm{NMAD}}$ is $0.034$ and $0.020$, for the \textit{HSC-like} and \textit{LSST-like} samples, respectively, as derived previously. 

The density contrast distributions for these \textit{Observational-like} samples are also shown in Figure \ref{hist_tot}, where the filled distributions in the second and fourth columns correspond to \textit{HSC-like} and \textit{LSST-like} samples, respectively. Additionally, we present in each panel a strict outlier fraction, $\eta$, using Equation \ref{eq:eta}: 
\begin{equation}
\raggedright
    \begin{split}
    \eta &= \frac{N_{\Delta z}[|\delta_z|/(1 + z_r) > \sigma_{\rm{NMAD}}]}{N_{\Delta z}}
\end{split}
\label{eq:eta}
\end{equation}
where $N_{\Delta z}$ is the number of photo-z selected galaxies within each redshift slab. This Figure also shows the galaxy surface density within the redshift slabs ($n_{\rm gal} = N_{\Delta z}/\pi $ deg$^{-2}$).
\subsubsection{Protocluster Overdensities}

Protoclusters are often found in observations as peaks in density contrast maps. To quantify their surface density in slabs of our survey emulations, we first extract all overdensity peaks in density contrast maps computed in the $200\times200$ pixels grid covering celestial coordinates $-1.0 < \Delta RA < 1.0$ [deg] and $-1.0 < \Delta Dec < 1.0$ [deg] (therefore, each pixel corresponds to $0.6$ arcmin). To identify the overdensity peaks, we have compared each pixel of these maps with the adjacent ones ($3 \times 3$ pixels matrix). If the central pixel presents the highest value, we select its position as an overdensity peak. Here, we did not apply any constraint in the overdensity value, $\delta_{\rm gal}$, of the peaks. However, we will define a detection threshold based on this value in the next section.

After, we analyze the surroundings of all protoclusters within each redshift slab. Our first step consists in identifying all peaks in protoclusters regions which are inside a radius equals to the bandwidth $\delta \theta$ centred at the position of a given structure. At the end of this step, we obtain two main samples: peaks in protocluster regions and those that are not. Since we expect protoclusters to be a collection of dark matter halos, they can contain more than one overdensity peak within $\delta \theta$. Nevertheless, given our choice of kernel bandwidth (effective radius of Coma-type protoclusters at $z_p$), this happens for less than $15\%$ of the structures in the entire sample. 
The next step consists of linking all peaks which are associated with protocluster regions to a given structure. In this case, we search for protoclusters inside the effective radius but centred at the position of the peaks. Then, the peak is associated to the most massive structure, if there are more than one protoclusters in this area.

The result of this exercise is also shown in Figure \ref{hist_tot}, where the black solid lines show the density contrast distributions of our \textit{HSC-like} and \textit{LSST-like} samples for the three types of clusters discussed here: Coma-type, Virgo-type, and Fornax-type. 
\begin{figure}
    \centering
    \includegraphics[width=\columnwidth]{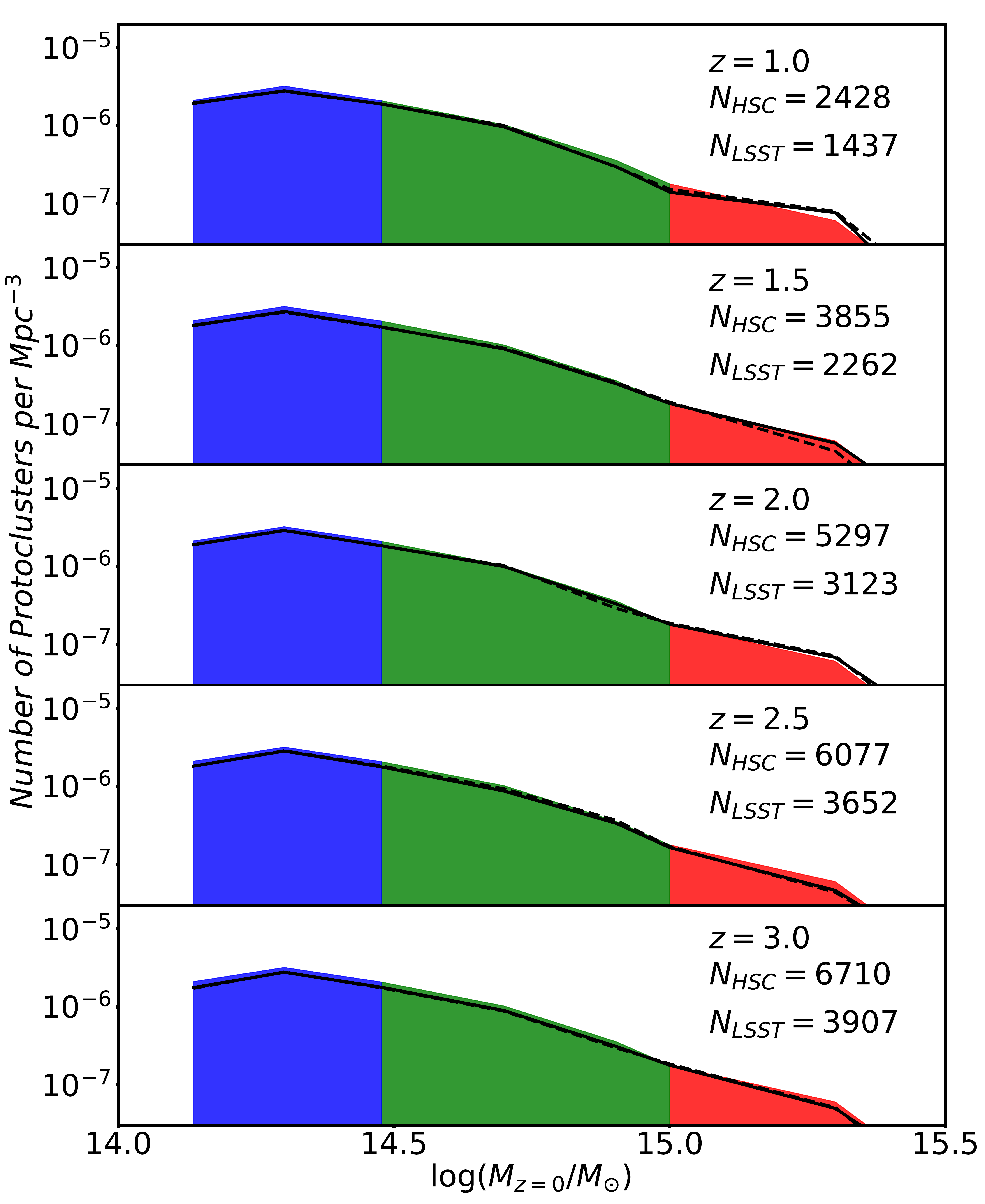}
    \caption{\small Descendant $z=0$ cluster mass ($M_{z=0}$) distribution of the selected protoclusters per Mpc$^{3}$ at $z_p=1.0$, $1.5$, $2.0$, $2.5$ and $3.0$, from top to bottom panels, respectively. Solid and dashed lines indicate the distribution for the \textit{HSC-like} and \textit{LSST-like} slabs, respectively, while filled areas represent those obtained from the $z=0$ Millennium simulation snapshot with different colours for Fornax, Virgo and Coma-type clusters. } 
    \label{cmass_dist}
\end{figure}

\begin{figure*}
    \centering
    \includegraphics[width=\textwidth]{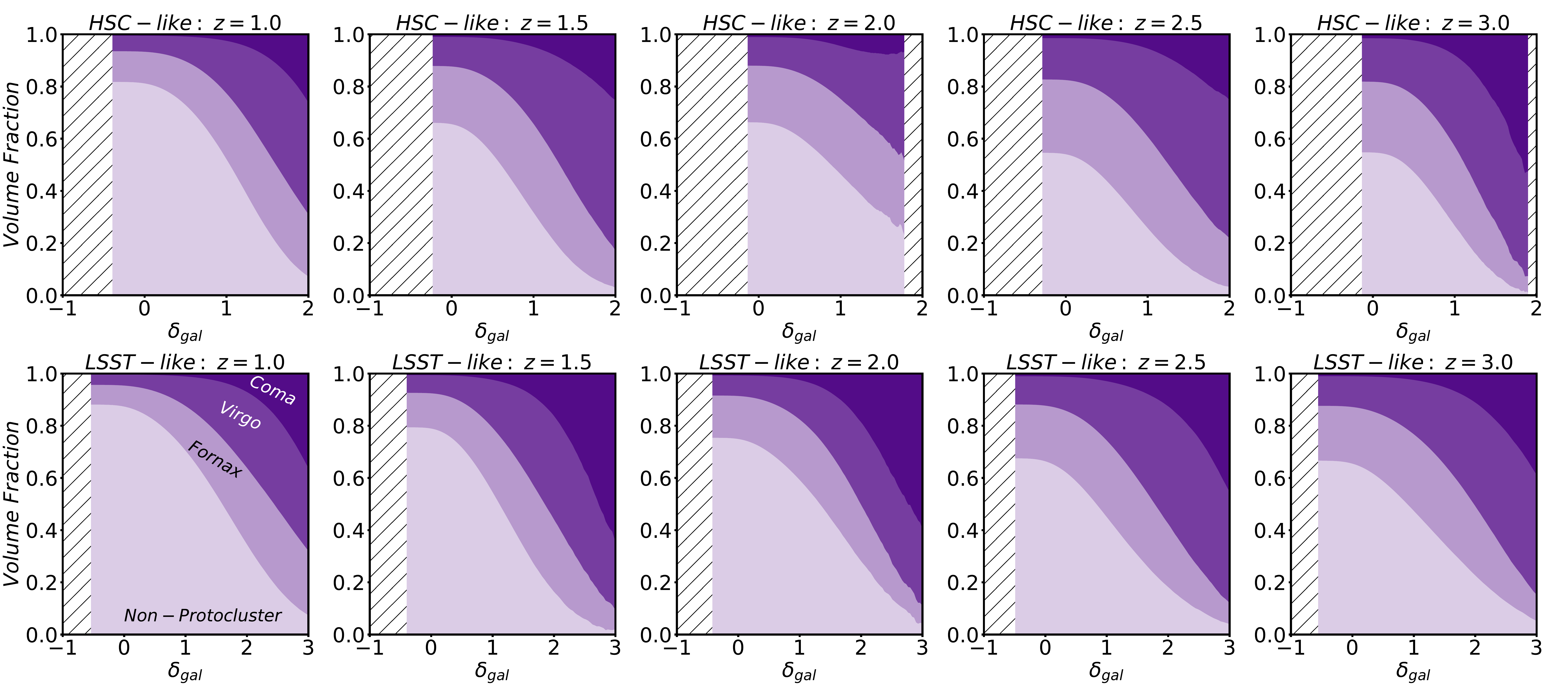}
    \caption{\small Probability of a peak with overdensity $\delta_{\rm gal}$ to be a non-protocluster (light purple), Fornax-type, Virgo-type, or Coma-type (dark purple) protocluster. These probabilities are based on overdensity estimations using a Gaussian KDE with bandwidth $\delta \theta$ (see Table \ref{tab:prob}), at $z = $ 1.0, 1.5, 2.0, 2.5 and 3.0, from left to the right. The first and second rows present results for the \textit{HSC-like} and \textit{LSST-like} samples, and both are magnitude limited at $i = 25.0$ mag. Hatched regions represent overdensity ranges without peaks.}
    \label{prob_pc}
\end{figure*}

\subsubsection{Protocluster Probabilities} \label{sec:pc_probs}

Now we investigate what is the probability that an overdensity peak with $\delta_{\rm gal}$ larger than a certain value is associated with a protocluster. As shown in Figure \ref{hist_tot}, protoclusters are found in regions with higher overdensity compared to the field, and we can use $\delta_{\rm gal}$ to control the purity of protocluster candidates selected by overdensities.

Similar to \citet{chiang14} work, but by analyzing the overdensities of the peaks instead of random positions, we classify our sample into four classes: (1) peaks in Coma, (2) Virgo, or (3) Fornax-type protocluster regions, and (4) those that are not associated with any structure of this type.

The inclusion of Coma-type progenitors in PCcones can generate biases when small volumes are analyzed (small field-of-view, or narrow redshift ranges), due to their low-density. To avoid this, we analyzed just the $80$ lightcones without placed structures of the type considered at $z_p$. We compare the descendant $z=0$ mass, $M_{z=0}$, distributions of the selected protoclusters within the redshift slabs of both emulated surveys with those from the $z=0$ snapshot of the Millennium simulation. We present this comparison normalized by the volume in Figure \ref{cmass_dist}. Since $\Delta z$ is not the same for HSC and LSST emulations, the number of selected protoclusters at a certain redshift slab is not the same, as depicted in this Figure.

The number of Coma progenitors is intrinsically small. For example, we found 49 and 31 of them at $z=1.0$ within \textit{HSC-like }\textit{LSST-like} slabs, respectively, compared with 1685 and 986 for Fornax-type protoclusters. Then, to avoid dealing with small samples, we modeled the overdensity distribution of the four different classes with a Gamma function, since the fitting presents a good agreement with the measurements. We use it to generate $N \times n_i$ random overdensity values associated with each class, where $N$ is a very large number and $n_i$ is the volume fraction of the $i$-th class.  
Finally, we estimate the probability to find a peak with overdensity $\delta_{\rm gal}$, associated with the $i$-th class, following Equation \ref{eq:prob}:
\begin{equation}
    P(\delta_{\rm gal}| i) = \frac{N_{p,i}(\geq \delta_{\rm gal})}{\sum N_{p,i}(\geq \delta_{\rm gal})},
    \label{eq:prob}
\end{equation}
where $N_{p,i}(\geq \delta_{\rm gal})$ is the number of peaks with overdensity higher than $\delta_{\rm gal}$, associated with the $i$-th type. Notice that the used value of $N_{p,i}$ comes from our sampling with the Gamma function.
We present these probabilities as a function of  $\delta_{\rm gal}$ in Figure \ref{prob_pc} for \textit{HSC-like} and \textit{LSST-like} mock surveys on the first and second rows, respectively, for redshifts $z = 1.0$, $1.5$, $2.0$, $2.5$ and $3.0$ (from left to right, respectively).

Figure \ref{prob_pc} shows, as expected, that by adopting higher values of $\delta_{\rm gal}$, we can achieve a higher probability of the selected peaks to be associated with a real structure. Based on this figure, we can establish criteria to classify protocluster candidates. We restrict our sample of peaks to achieve a 70$\%$ confidence level that they are genuine protoclusters. To obtain this overdensity lower limit, we find the value when the volume fraction of non-protoclusters drops to 0.3, which is equivalent to the desired detection accuracy. 
On the other hand, by conditioning the sample with respect to $\delta_{\rm gal}$, we limit the total number of detected structures that are real, i.e., the completeness decreases. Additionally, as Figure \ref{prob_pc} shows, the protocluster probabilities are descendant-mass dependent. Therefore, the completeness is different for Coma, Virgo, and Fornax-type protoclusters.  We summarize our results of detection completeness for the adopted criteria in Table \ref{tab:prob} for \textit{HSC-like} and \textit{LSST-like} samples for different redshift slabs. 

\begin{table*}
	\centering
	\caption{\small Predicted galaxy overdensity, $\delta_{\rm gal}$, required to have $70 \%$ of probability of being a real protocluster for the \textit{HSC-like} and \textit{LSST-like} samples, which are limited to galaxies brighter than $i = 25.0$ mag. We also show the expected completeness associated to this criteria for the full sample ($C_{\rm all}$), Coma ($C_{\rm C}$), Virgo ($C_{\rm V}$), and Fornax type ($C_{\rm F}$) protoclusters, as well as the kernel bandwidth $\delta \theta$.}
	\label{tab:prob}
	\begin{adjustbox}{max width=\textwidth}
	\begin{tabular}{ccccccccccccccc} % four columns, alignment for each
	\hline
	  & \hspace{0.1cm}&\hspace{0.1cm}&\multicolumn{5}{c}{\hspace{2.0cm}\textit{HSC-like}} &\hspace{0.1cm}& \multicolumn{5}{c}{\hspace{2.0cm}\textit{LSST-like}} \\
	 		%\hline
       $z$&\hspace{0.05cm}&$\delta \theta \ [\mathrm{arcmin}] $&\hspace{0.05cm}&$\delta_{\rm gal}$&$C_{\rm all} \ [\%]$&$C_{\rm C}\ [\%]$&$C_{\rm V}\ [\%]$&$C_{\rm F} \ [\%]$& \hspace{0.05cm}&$\delta_{\rm gal}$&$C_{\rm all} \ [\%]$&$C_{\rm C}\ [\%]$&$C_{\rm V}\ [\%]$&$C_{\rm F}\ [\%]$ \\
         \hline
1.0 &\hspace{0.05cm}& 3.53 &\hspace{0.05cm} & 1.39 & 11.45 & 61.22 & 17.72 & 7.42 &\hspace{0.05cm}& 2.08 & 8.28 & 54.84 & 11.67 & 5.38 \\
1.5 &\hspace{0.05cm}& 4.21 &\hspace{0.05cm} & 1.01 & 13.00 & 48.65 & 19.87 & 9.20 &\hspace{0.05cm}& 1.54 & 5.17 & 41.46 & 7.26 & 3.37 \\
2.0 &\hspace{0.05cm}& 4.20 &\hspace{0.05cm} & 1.48 & 0.21 & 0.92 & 0.34 & 0.13 &\hspace{0.05cm}& 2.02 & 0.93 & 8.57 & 1.32 & 0.54 \\
2.5 &\hspace{0.05cm}& 4.31 &\hspace{0.05cm} & 0.89 & 18.18 & 58.33 & 29.16 & 12.90 &\hspace{0.05cm}& 1.49 & 13.94 & 63.64 & 25.37 & 8.11 \\
3.0 &\hspace{0.05cm}& 4.33 &\hspace{0.05cm} & 0.93 & 6.23 & 29.84 & 10.56 & 3.90 &\hspace{0.05cm}& 1.72 & 10.06 & 51.95 & 18.06 & 5.77 \\

		\hline
	\end{tabular}
	\end{adjustbox}
\end{table*}

If we adopt other values of $\delta \theta$ as kernel bandwidth, a different $\delta_{\rm gal}$ threshold has to be applied for obtaining a protocluster candidates sample at a 70\% confidence level. For example, smaller values of $\delta \theta$ produce more contrast between the field and dense environments (the amplitude of overdensities increases). However, the number of spurious detections also increases. Then, we need to impose a more strict overdensity threshold to achieve 70\% purity. In general, the completeness could vary about $\sim 7\%$ with respect to the actual values (depending on the redshift). Anyway, this is an issue that deserves further investigation. 

In Appendix \ref{sec:cluster_dec} we focus on the detectability of the 20 protoclusters placed at $z=1.0$, 1.5, 2.0, 2.5, and 3.0 for HSC-SSP and LSST mock configuration \citep[see also][]{vicentin20}. There, we also discuss particular cases of non-detected structures.

\subsection{Analysis of the CFHTLS} \label{sec:cfht-like}
\begin{figure}
    \centering
    \includegraphics[width=\columnwidth]{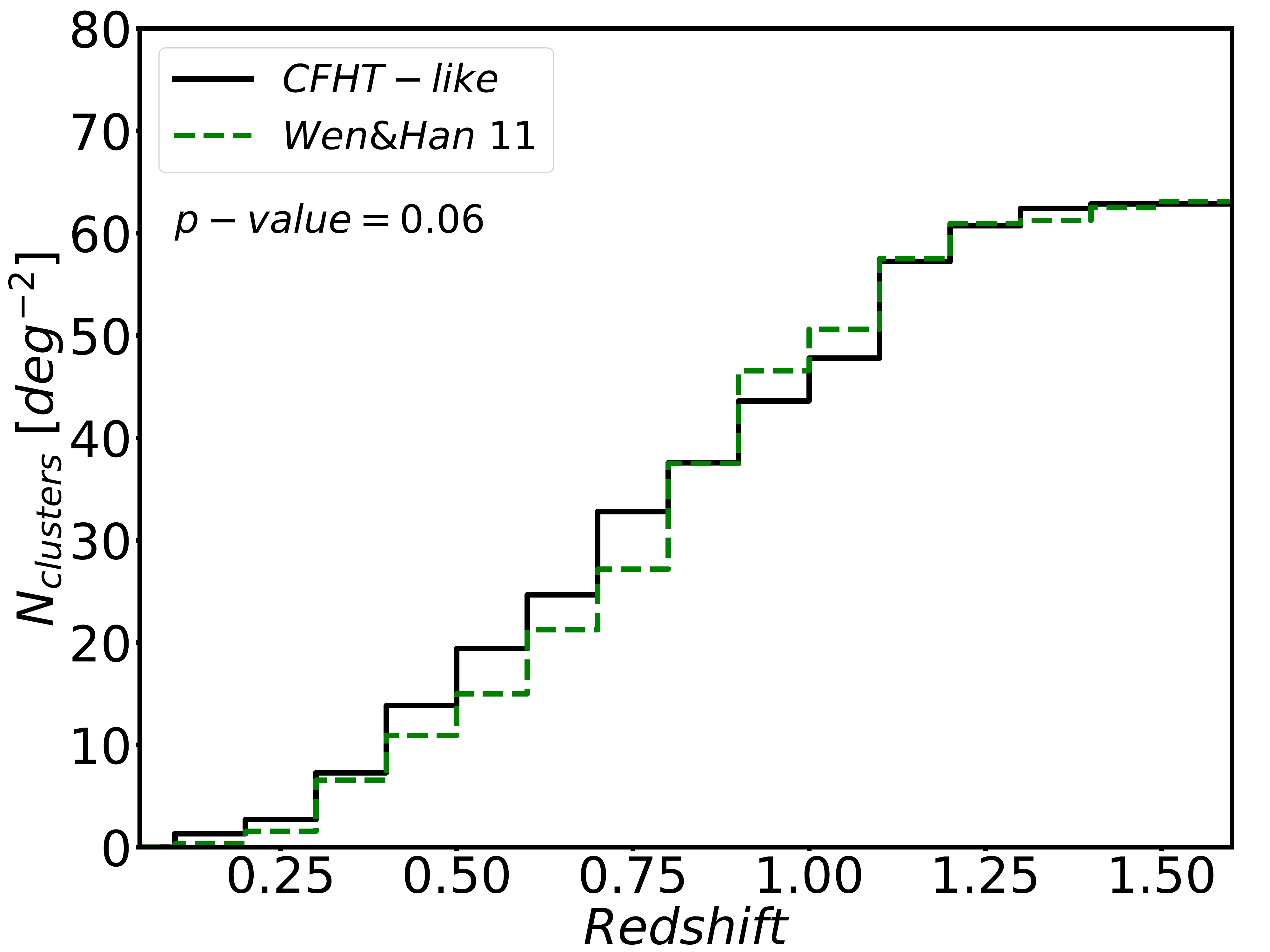}
    \caption{\small Cumulative cluster per deg$^2$, $N_{\mathrm{clusters}}$ per redshift bin, from the cluster detection of \citet{wen11} (green dashed line) for the Deep CFHTLS. Our results from the deep \textit{CFHT-like} sample are shown as a black continuous line.}
    \label{cfht_dist}
\end{figure}
We have been focusing on protocluster detection at specific redshifts. In this section, we emulate continuous cluster detection at $z \leq 1.5$, using the \textit{CFHT-like} sample that simulate the Deep Canada France Hawaii Telescope Legacy Survey (CFHTLS; \url{http://www.cfht.hawaii.edu/Science/CFHTLS/}). To do this, we compute the overdensity maps at 29 different overlapping redshift slabs centred at $z = 0.1$ to $1.5$ and spaced by redshift intervals of $0.05$. We have selected all galaxies with photo-z within $z_i \pm \Delta z = z_i \pm \sigma_{\rm{NMAD}}^{CFHT} \times (1+z)$, where $ \sigma_{\rm{NMAD}}^{CFHT} = 0.027$ and $z_i$ is the centre of each slab. We use in this case, for simplicity, a Gaussian kernel with bandwidth $\delta \theta = 1.0$ Mpc. Therefore, the density field here is described by a parallelepiped composed by 29 slabs. We apply a procedure similar to that described earlier to extract overdensity peaks, but now using a $3\times3\times3$ pixels  analyser cube (corresponding to 0.6 $\times$ 0.6 in arcmin and 0.05 in redshift) that runs over all the 29 maps data cube.  Hence, if the central pixel of the analyzer cube is a maximum, then we register its position. 

We repeat this procedure for 6 PCcones without putting any {\it a priori} structures within the redshift interval of interest, because it could add bias to the cluster distribution.

Finally, we define as cluster candidates all overdensity peaks with $\delta_{\rm gal} \geq 2.0$. This value is the same density contrast threshold used by \citet{durret11}. It corresponds to $\rm{median}(\delta_{\rm gal}) + \rm{std}(\delta_{\rm gal})$, where $\rm{median}(\delta_{\rm gal})$ and $\rm{std}(\delta_{\rm gal})$ are the median and the standard deviation of $\delta_{\rm gal}$, respectively. However, since we are using a fixed bandwidth in Mpc, maps at high redshift present a large number of overdense peaks, because $\delta \theta$ decreases in angular coordinates,  the number of galaxies decreases, and these two factors tend to amplify the density contrast. Moreover, these peaks can enclose just a few galaxies within a radius of 1 Mpc. To avoid these statistical fluctuations we establish a minimum number of galaxies  inside a radius of 1 Mpc, $N_{min} = 17$, which represents the median number of galaxies of overdensity peaks with $\delta_{\rm gal} \geq 2.0$. 
The cumulative redshift distribution for the mock cluster candidates is shown in Figure \ref{cfht_dist} (black solid line). It also presents the cluster distribution obtained by \citet{wen11} (green dashed line) for the Deep CFHTLS. This survey has a 3.2 deg$^2$ field with $5\sigma$ magnitude limits presented in Table \ref{tab:ground}. \citet{ilbert06} and \citet{coupon09} obtained redshift uncertainties $\sigma_{\Delta z}/(1+z) = 0.029$ for galaxies with $i < 24.0$ for this field. Following \citet{wen11}, we adopted here  the same magnitude limit. We have performed a \textit{Kolmogorov-Smirnov} test to verify whether the  \citet{wen11} sample and our simulated cluster distribution are consistent with each other. We obtained a $p-value=6.1\%$, indicating that both distributions are somehow similar. We can also obtain a cluster redshift distribution similar to the \citet{wen11} catalogs with other kernel bandwidths $\delta \theta$. For example, if  $\delta \theta = 1.5$ Mpc, we obtain a $p-value=13.0\%$ when we limit our sample to $\delta_{\rm gal} \leq 1.30$ and $N_{\rm min} = 19$. 

On the other hand, with more restrictive values of $\delta_{\rm gal}$ (or $N_{\rm min}$), we obtain a catalog of cluster candidates with higher purity. The completeness, however, tends to drop, and the resulting redshift distributions are no longer consistent with \citet{wen11}.

\section{Discussion}

The detection of structures at high redshifts, such as galaxy clusters or protoclusters, is critically affected by observational constraints. We have explored in this work the impact of magnitude constraints and photometric redshift estimations on  the analysis of the galaxy density field derived from photometric surveys by using a new implementation of mock catalogs that we call PCcones. It is a tool that allows to emulate galaxy surveys, taking into account their observational constraints.  The particularity of PCcones is that a selected object is placed at the centre of each mock. This helps the study of rare structures, such as massive protoclusters, in particular how observational constraints affect their detection and estimation of physical properties.

Additionally, we have adopted here a different procedure to obtain apparent magnitudes for the Millennium Lightcone that uses the post-processing technique of \citet{shamshiri15}. This is done by ascribing, to all mock galaxies, a SED computed from \texttt{L-GALAXIES} \citep{henriques15} star formation histories. 

In Section \ref{sec:photo_z} we have presented results on photometric redshift estimations for PCcones and the Millennium Lightcone, using the \texttt{Le Phare} SED fitting algorithm. We notice that our approach improves these estimations due to our method to obtain observer-frame magnitudes. For instance, the Millennium Lightcone presents a clear bias at $z \lesssim 1.0$, even considering that both mocks were computed using the same SAM. Recently, \citet{laigle19} presented predictions for photometric redshifts and physical properties of galaxies for the LSST survey by using mock catalogs based on hydrodynamical simulations. \citet{laigle19} also attributed a SED to each simulated galaxy to obtain apparent magnitudes. Their results for photo-zs are similar to ours. For example, for galaxies in the $i$ band bin $22 < i \leq 23$ mag, we obtain  $\sigma_{\rm{NMAD}} = 0.017$ and an outlier fraction $f_{\rm{outliers}} = 0.1 \%$, which are exactly the same values obtained by \citet{laigle19}. However, for deeper $i$ band magnitude bins, our photo-z predictions are more accurate. These differences can be due be due to the differences in the simulations and in the IGM absorption model (they applied the IGM absorption to the galaxy SEDs according with the gas distribution in the simulated IGM). We have also shown, in Section \ref{sec:photo_zs}, that our Deep \textit{CFHT-like} mock presents a difference from the observational results of \citet{ilbert06} of only about $2\%$. Our results in that section indicate that PCcones reproduce reliably observed photo-z estimates without the implementation of additional codes to obtain mock magnitudes such as \texttt{PhotReal} \citep{ascaso15}. We can accurately emulate surveys/observations by adding realistic noise to PCcones magnitudes. This is useful to test different approaches to photo-z estimation and structure detection.

It is a common practice in photometric surveys to look for structures in redshift slabs and, in this case, the slab width plays a critical role. \citet{chiang1} has examined the impact of redshift uncertainties on protocluster overdensities, noticing that with a slab width $\Delta z = 0.15$, many random regions could be spurious candidates. 

We notice that, as expected, the distribution of the density contrast, $\delta_{\rm gal}$, of the field (filled histogram in Figure \ref{hist_tot}) is different between the \textit{Pure Simulation} and \textit{Observational-like} samples, mainly when the number of galaxies in the field is low or the outlier fraction $\eta$ is high. 

In Figure \ref{prob_pc}, we present the probability of an overdensity peak in our maps be a real structure. This figure helps us define a detection threshold to classify regions as protocluster candidates based on their overdensity. We have analyzed protocluster detectability to achieve $70\%$ of purity. To find this threshold, we estimate at what $\delta_{\rm gal}$ value the non-protocluster distribution drops to $0.3$. Similarly, other criteria can be adopted, for example, to reach $60\%$ or $80 \%$ of selection accuracy.

We also present in Figure \ref{prob_pc} the volume fraction of different cluster type progenitors as a function of the overdensity $\delta_{\rm gal}$. As expected, Coma-type protoclusters are associated with high overdensities. However, we found a non-negligible fraction of Virgo and Fornax-type protoclusters with very high overdensities as well. This is due to projection effects, where different independent structures in the same area amplify the overdensities of low mass protoclusters. This effect is more noticeable in the \textit{HSC-like} sample because the redshift slab is larger. We found that the mean number of protoclusters within the effective radius matched with Virgo-types is $\gtrsim 2.3$ at the redshifts where this trend is stronger.

Table \ref{tab:prob} shows that the completeness at the $70 \%$  confidence level for the \textit{HSC-like} sample is higher than those obtained for the \textit{LSST-like} one, in almost all redshifts. This does not imply, however, that structure detection for the wide HSC-SSP is better than the expected for LSST, whose photo-z estimations are more accurate. If we reanalyze this sample with the same slab width used for the \textit{HSC-like} one, we find that  $C_{\rm all }$ for \textit{LSST-like} is in general higher or equal to that obtained for \textit{HSC-like}. This occurs because, for higher redshift slabs, there are more structures in the same area (see Figure \ref{cmass_dist}), and since the number of overdensity peaks in both fields is similar, the chance to match protoclusters with overdensity peaks increases for the \textit{HSC-like} sample.

Figure \ref{prob_pc} and mainly  Table \ref{tab:prob} show evidence that structure detectability presents a strong dependence with redshift, mostly due to the variation in the accuracy of photo-zs. This suggests that detection criteria should be established as a function of redshift. Notice that the completeness of the protocluster sample is lower when $\eta$ is high ($\gtrsim 60\%$). This occurs at $z = 2.0$ and 3.0 for the \textit{HSC-like} sample, where the achieved completeness is 0.21\% and 6.23\%, respectively. It also happens at $z=2.0$ in \textit{LSST-like} emulation, where the expected completeness is 0.93\%. Certainly, protocluster detection would be improved by adding available spectroscopic information, and we will analyze and quantify its impact in terms of purity and completeness in a future work.

The magnitude limit plays an important role in overdensity estimations due to the \textit{Malmquist-bias}, which affects galaxy completeness. The construction of density contrast maps with low galaxy completeness leads to shot-noise, where a small set of galaxies generates overdense regions \citep[e.g.][]{vicentin20}. However, this does not happen in our case, given the imposed magnitude limit ($i = 25.0$ mag). 

We can compare our results with  similar works. \citet{chiang14} obtained a sample of protocluster candidates by using photometric redshifts from the COSMOS/UltraVista catalog \citep{muzzin13}, whose photo-z uncertainty is $\sigma_{z} = 0.025(1+z)$. Their  detection threshold is the mean overdensity of Coma-type protoclusters, based on lightcones. They achieved a purity of $\sim 70\%$, and completeness of 9\%, 7\%, 17\%, and 50\% for all, Fornax, Virgo, and Coma-type structures, respectively, at $1.6 < z < 3.1$. Considering redshift slabs at $1.5 \leq z \leq 3.0$, we obtain, on average, completeness of $\sim$ 7\%, 4\%, 13\%, and 40\% for the same classes, respectively, in the \textit{LSST-like} sample.         
\citet{ando20} constructed another catalog of protocluster candidates by searching for pairs of massive galaxies in the COSMOS2015 data \citep{laigle19} at $z \sim 2.0$.
Given their selection criteria, they estimated that 54\% of the observed galaxy pairs are real and that 63\% of them reside in protocluster regions ($M_{z=0} > 10^{14} \ M_{\odot}$). From simulations, they estimated a completeness of 23\%, 16\%, 52\%, and 100\% for the whole, Fornax, Virgo, and Coma-type protocluster samples, respectively. At the same confidence level, for the \textit{HSC-like} sample, our completenesses are $\sim31\%$, 26\%, 42\%, and 64\%.

Despite all these factors that affect protocluster detection in photometric surveys and the simplicity of our method, we have obtained acceptable results for our HSC-SSP and LSST emulations. For example, we can achieve a completeness ranging from $\sim 12\%-18\%$ for a 70\% confidence level at $z_p = 1.0$, 1.5 and 2.5 for the \textit{HSC-like} sample. In the case of the \textit{LSST-like} sample, our best results are at $z_p = 2.5$ and 3.0; here, we can recover $13\%-18\%$ of all protoclusters with 70\% of purity in a narrower redshift slab. To reach these results, we have to apply a threshold in the overdensities (see $\delta_{\rm gal}$ in Table \ref{tab:prob}) of the peaks in the fields at the considered redshifts. With our definition of candidates, we can detect, on average, $\sim39\%$ and 43\% of the Coma-type progenitors for the \textit{HSC-like} and \textit{LSST-like} samples, respectively at $1.0 \leq z \leq 3.0$. Given the expected number of Coma-type protoclusters in this redshift interval in each survey (sky coverage of 1,400 deg$^2$ and 20,000 deg$^2$, for HSC-SSP and LSST, respectively), we should be able, with our approach, to detect $\sim$3033 and $\sim$43,329 progenitors of the most massive clusters in the full HSC-SSP and LSST areas, respectively.

\section{Summary}

In this work we have introduced a new emulation tool that can implement observational constraints in simulations of galaxy surveys. It was applied here to analyze the level of significance of overdensity estimations associated to the search of protoclusters of galaxies. Our main results are as follows:

\begin{enumerate}
    \item We have developed a procedure that allows to build a mock lightcone that contains some desired structure at the centre of the line-of-sight and at a certain redshift of interest. These PCcones have been constructed by redefining the mock zero-point (i.e., $z=0$). This is important to place correctly the centre of masses of a protocluster at a given redshift $z_p$ (see Section \ref{sec:lightcone_space}).  
    
    \item We have estimated mock apparent magnitudes at a \textit{post-processing} stage, using the SFH arrays generated by the \texttt{L-GALAXIES} semi-analytic model of \citet{henriques15}. Galaxy SEDs have allowed us to obtain reliable observer-frame magnitudes, without using magnitude interpolations or K- corrections (see Section \ref{sec:galaxy_sed}). 

    \item We have estimated photometric redshifts in mock surveys (see Section \ref{sec:photo_z}), using \texttt{Le-Phare}. To do this, we have created \textit{Observational-like} mocks, emulating the Deep CFHTLS (\textit{CFHT-like}), the wide layer of HSC-SSP (\textit{HSC-like}), and the 10 years forecast of LSST (\textit{LSST-like}), by applying observational constraints on the photometric bands of each of these surveys, accordingly to their $5\sigma$ magnitude limits.
    
    \item The comparison of our mock photometric redshifts with their observational counterpart for the  \textit{CFHT-like} and \textit{HSC-like} samples gives satisfactory results (see Figure \ref{photo-z_surveys}). The comparison of our photo-z predictions are consistent with other LSST photo-z forecasts. Additionally, we have estimated photo-zs using the Millennium Lightcone from \citet{henriques15} for the \textit{CFHT-like} sample, finding that the Millenium photometric redshifts present a clear bias at $z \lesssim 1.0$ (see Figure \ref{photo-z}), contrarily to what we have obtained. We attribute this difference to the procedure we adopted to compute the apparent magnitudes. 

    \item We have constructed galaxy density contrast maps for the PCcones at five different redshifts ($z=$ 1.0, 1.5, 2.0, 2.5 and 3.0), using a two-dimensional Gaussian kernel (Figure \ref{overdensity_panels} shows an example for a protocluster at $z=1.0$). We have focused on overdensity estimations for two types of samples: the \textit{Observational-like} sample, composed by the \textit{HSC-like} and \textit{LSST-like} mock surveys, and, for comparison, the \textit{Pure Simulation} sample, without any observational/magnitude constraints. For both samples, we have obtained density maps in slabs using our estimated photometric redshifts.
    
    \item For each density contrast map we extracted all overdensity peaks within a redshift slab and investigated their association to protoclusters. This allowed us to estimate the probability of protocluster detection at a given density contrast $\delta_{\rm gal}$, as well as the completeness of this sample at a $70\%$ confidence level (probability to be real protoclusters). 
    
    \item Our main results are summarized in Table \ref{tab:prob}. For the HSC-SSP emulation, we expect to recover: $\sim 48\%-61\%$, $\sim 17\%-29\%$, and $\sim 7\%-12\%$ of Coma-, Virgo- and Fornax-type protoclusters, respectively, at $z_p= 1.0$, 1.5, and 2.5. For the 10-year forecast of the LSST, within a narrower redshift slab, these numbers are: $\sim 51\%-62\%$, $\sim 11\%-25\%$, and $\sim 5\%-8\%$ at $z_p = 1.0$, 2.5 and 3.0.
    
    \item In some cases, the combination of observational constraints and photo-z uncertainties affect the detection of structures critically. This happens at $z_p= 2.0$ for both emulated samples. We found that the completeness of all protoclusters is $\sim 1\%$, given our selection criterion, for this redshift. These results suggest that wide-field spectroscopy will be needed if we want to achieve anything remotely approaching a complete protocluster sample.
    
    \item Table \ref{tab:prob} shows that structure detectability (the $\delta_{\rm gal}$ required to achieve 70\% of purity) changes drastically with redshift, mostly due to the decreasing accuracy of photo-zs. This suggests that the structure selection criteria should be established as a function of redshift.

    \item Finally, we have emulated cluster detection in a \textit{CFHT-like} sample, obtaining  cluster distributions similar to actual observations. 
\end{enumerate}

\section*{Acknowledgements}

We thank our referee, Brian Lemaux, whose positive comments and suggestions helped us to improve the paper. Also, We thank Bruno M. B. Henriques for helping us to obtain the post-processing magnitudes, and also to make publicly available the \texttt{L-GALAXIES} SAM, Gerald Lemson for giving us access to the Virgo-Millennium database, Melissa Graham for her explanation on the procedure to estimate observation-like magnitudes, and, finally, Doris Stoppacher, Eduardo Cypriano, Gast\~ao Lima-Neto and Cristina Furlanetto for useful discussions and comments. P.A.-A. thanks the Conselho Nacional de Desenvolvimento Cient\'ifico  e Tecnol\'ogico (CNPq-Brasil) for supporting his MSc scholarship (project 133350/2018-5). M.C.V. acknowledges support from CAPES/FAPESP (project 2018/01469-6). L.S.J. acknowledge support from FAPESP (project 2019/10923-5) and CNPq (304819/2017-4). R.A.O. was supported by grants from CNPq (302981/2019-5) and FAPESP (2018/02444-7).
The Millennium Simulation databases used in this paper and the web application providing online access to them were constructed as part of the activities German Astrophysical Virtual Observatory (GAVO).

\section*{Data Availability}

The PCcones are catalogs of 100 independent lightcones (20 protoclusters placed at $z=1.0$, 1.5, 2.0, 2.5 and 3.0). To access this data contact Pablo Araya-Araya (e-mail: paraya-araya@usp.br), asking for specific mocks or the full dataset. Also, some scripts such as those to obtain observational-like magnitudes, overdensities measurements for the \textit{HSC-like} and \textit{LSST-like} samples and for cluster detection in the \textit{CFHT-like} case will be stored in \url{https://github.com/Pabl1to/PCcones_codes}.

%\bibliographystyle{mnras}
\bibliographystyle{mnras}
\bibliography{bibliography}

%%%%%%%%%%%%%%%%%%%%%%%%%%%%%%%%%%%%%%%%%%%%%%%%%%

%%%%%%%%%%%%%%%%% APPENDICES %%%%%%%%%%%%%%%%%%%%%

\appendix

\section{Photo-z comparison with Millennium Lightcone} \label{sec:comparison}
We have performed a comparison of two different methods, that of \citet{henriques15} and ours, to obtain apparent magnitudes from the Millennium Lightcone using the same galaxies in both cases. 

Magnitude distributions are similar to each other, as Figure \ref{g_count_mag} shows. However, we found that the median of the absolute difference between galaxy magnitudes, $median(|m_{\rm our} - m_{\rm ml}|)$ are $\sim 0.23$, $0.18$, $0.19$, $0.21$ and $0.24$ $[mag]$ for the $u$, $g$, $r$, $i$ and $z$ photometric bands, respectively. 

\begin{figure}
    \centering
    \includegraphics[width=\columnwidth]{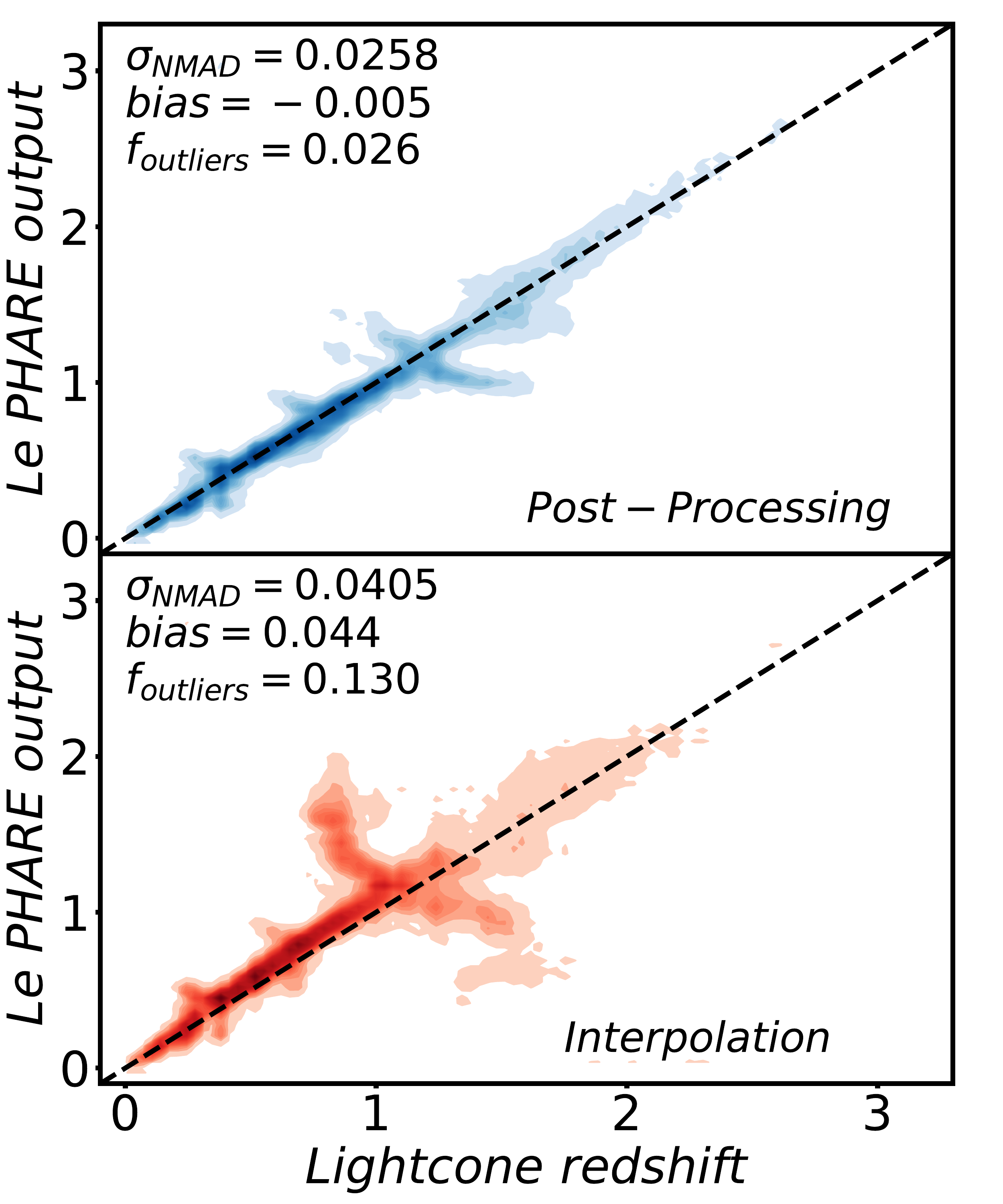}
    \caption{Photometric redshift estimation using  \texttt{Le Phare}. We compare results from  the Millennium Lightcone using \textit{post-processing} apparent magnitudes (blue contours, first panel) with those obtained with interpolated magnitudes (red contours, second panel). Photo-zs were estimated with the photometric constraints of the deep CFHTLS.}
    \label{fig:comparison_mill}
\end{figure}

After, we generated observational-like catalogs, emulating again the Deep CFHTLS and then we run \texttt{Le Phare}. Since we obtain observational-like magnitudes as Gaussian random values with median mock magnitudes and standard deviation equal to the expected error, the galaxies are not anymore the same in both catalogs. However, we find that $\sim 86 \%$ of the galaxies are in both observational-like catalogs. Figure \ref{fig:comparison_mill} shows galaxies that were found in both datasets. This figure is very similar to Figure \ref{photo-z} and indicates that the difference between the quality of our estimated photometric redshifts and those obtained from the Millennium Lightcone is due to the different approaches in estimating the magnitudes of each catalogue.
\section{Cluster detectability} \label{sec:cluster_dec}

In this section, we focus on the 20  protoclusters placed at $z=1.0$, 1.5, 2.0, 2.5, and 3.0 (Section \ref{sec:pcones}). We summarize the overdensity measurements of these structures in Table \ref{tab:protoc_obs}, for the two surveys. Protoclusters that do not appear in the table are those  without an overdensity peak within a projected distance $R_e(z)$ of their centres. For the HSC-SSP simulation we have detected, at these redshifts, 16, 14, 8, 14, and 11 protoclusters, respectively. For the LSST forecast, the corresponding numbers are 18, 15, 12, 14 and 13, respectively. 

\begin{figure}
    \centering
    \includegraphics[width=\columnwidth]{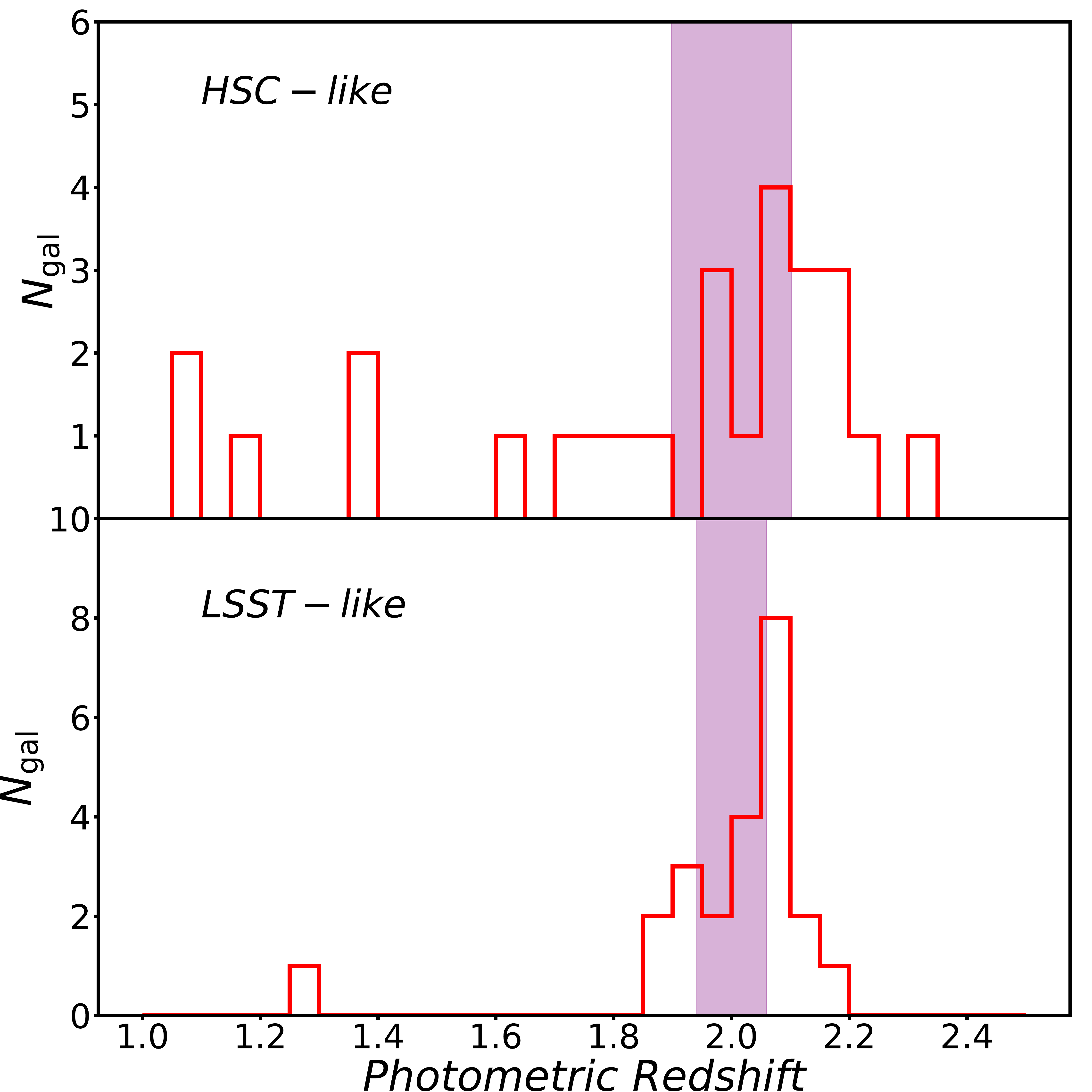}
    \caption{Photometric redshifts distribution (\textit{HSC-like} and \textit{LSST-like} photo-zs in the top and bottom panels, respectively) of the Coma-type protocluster with $\log{(M_{z=0}/M_{\odot})} = 15.16$ which was not detected. The shaded region represents the redshift slab.}
    \label{mem_zdist}
\end{figure}

Considering  Coma-type protoclusters, we recovered at least 4 out of 6 systems; an exception is for $z=2.0$  for the \textit{HSC-like} case, where only  2 out of 6 were detected. This is due mainly to photo-z uncertainties, since at $z=2.0$, many protocluster members (independently of the cluster type) have a measured photo-z outside the redshift slab for both simulated surveys. This can be seen in Figure \ref{mem_zdist}, where we show the photometric redshift distribution of member galaxies of a Coma-like non-detected protocluster ($\log{(M_{z=0}/M_{\odot})} = 15.16$) for both survey emulations. Therefore, photometric redshift uncertainties can dilute real overdensities. This is not the case for the non-detected Coma-type at $z=1.0$, however. This particular structure is marked with a red circle in Figure \ref{overdensity_panels}, and it is, in fact, in a dense environment, since it concentrates $\sim 77$ photo-z selected members in the main dark matter halo within a radius of 2.0 arcmin and is $\sim 3.3$ times denser than the field. Nevertheless, the detected overdensity peak suffered a shift due to the projection of galaxies in secondary halos, some of them members of a massive group ($\log{(M_{z=0}/M_{\odot})= 13.70}$). In conclusion, both photometric redshift uncertainties and projection effects may preclude even the detection of rich structures.

\begin{table*}
	\centering
	\caption{\small Detected overdensities of the 20 protoclusters for the \textit{HSC-like} and \textit{LSST-like} samples at $z=1.0$, 1.5, 2.0, 2.5 and 3.0.}
	\label{tab:protoc_obs}
	\begin{adjustbox}{width=\textwidth}
	\begin{tabular}{c|c|ccccc|ccccc} % four columns, alignment for each
	\hline
	  \multicolumn{1}{c|}{}&\multicolumn{1}{c|}{}& \multicolumn{5}{c|}{$\delta_{\rm gal}$ (\textit{HSC-like})} & \multicolumn{5}{c}{$\delta_{\rm gal}$ (\textit{LSST-like})} \\
	 		%\hline
       Descendant Type &$\log{(M_{z=0}/M_{\odot})}$&$z=1.0$&$z=1.5$&$z=2.0$&$z=2.5$&$z=3.0$&$z=1.0$&$z=1.5$&$z=2.0$&$z=2.5$&$z=3.0$\\
         \hline
\multirow{8}{4em}{Fornax}&14.14 & 0.51 & 0.43 & --- & 0.58 & --- & 1.15 & --- & --- & 0.75 & --- \\
&14.17 & 1.37 & --- & --- & 1.45 & 0.53 & 1.77 & 1.17 & 1.45 & --- & 2.32 \\
&14.20 & 1.20 & --- & --- & --- & 1.01 & 1.54 & 1.62 & --- & --- & 1.18 \\
&14.23 & --- & 0.67 & --- & 0.72 & --- & 0.76 & 0.77 & 0.97 & 1.22 & 1.46 \\
&14.27 & 0.88 & 0.86 & --- & 1.02 & --- & 1.46 & 0.86 & --- & --- & 1.43 \\
&14.31 & 1.22 & 0.93 & --- & --- & 0.61 & 1.77 & 0.75 & 0.64 & 0.85 & 1.81 \\
&14.36 & --- & 0.53 & 0.72 & 1.01 & --- & 0.98 & 0.42 & --- & 0.95 & --- \\
&14.42 & 0.50 & --- & 0.34 & --- & 0.87 & 0.34 & --- & 0.93 & --- & --- \\
\hline
\multirow{6}{4em}{Virgo}&14.49 & --- & --- & --- & 1.27 & --- & --- & 1.10 & --- & --- & --- \\
& 14.58 & 1.33 & 0.98 & 0.66 & 0.56 & --- & 1.43 & --- & 1.09 & 1.27 & 1.12 \\
& 14.73 & 1.09 & 0.83 & 0.75 & 0.76 & --- & 1.69 & 1.20 & --- & 1.20 & 1.81 \\
& 14.73 & 1.31 & 1.07 & 0.74 & 1.31 & 0.66 & 1.55 & 1.44 & 1.17 & 1.62 & 1.64 \\
& 14.82 & 1.42 & 1.17 & 0.63 & 1.03 & 0.84 & 2.16 & 1.10 & --- & 1.79 & 1.76 \\
& 14.94 & 1.34 & 0.91 & --- & --- & --- & 2.44 & 0.81 & 0.47 & --- & --- \\
\hline
\multirow{6}{4em}{Coma}& 15.03 & 1.31 & 0.91 & --- & 0.90 & 0.46 & 2.55 & --- & 0.67 & 1.48 & 1.50 \\
&15.05 & 2.22 & 1.90 & --- & --- & --- & 3.02 & 2.21 & 1.66 & 2.18 & --- \\
&15.07 & 1.10 & --- & --- & --- & 1.00 & 1.87 & --- & 1.34 & 1.97 & 1.82 \\
&15.16 & 1.09 & --- & --- & 1.30 & 1.03 & 2.12 & 1.15 & --- & 2.49 & 2.04 \\
&15.26 & --- & 1.55 & 0.87 & 1.51 & 0.85 & --- & 1.67 & 1.65 & 1.74 & 2.29 \\
&15.36 & 2.61 & 1.59 & 1.09 & 1.12 & 0.74 & 4.40 & 1.64 & 1.60 & 2.22 & --- \\
		\hline
	\end{tabular}
	\end{adjustbox}
\end{table*}
%%%%%%%%%%%%%%%%%%%%%%%%%%%%%%%%%%%%%%%%%%%%%%%%%%

% Don't change these lines
\bsp	% typesetting comment
\label{lastpage}
\end{document}